%% file: main.tex
\documentclass[a4paper,fleqn]{cas-dc}

\usepackage{algorithm}
\usepackage{dsfont}
\usepackage{lineno}
\usepackage[utf8]{inputenc}
\usepackage[T1]{fontenc}
\usepackage[flushleft]{threeparttable}
\usepackage{xspace}
\usepackage{amssymb}
\usepackage{xcolor}
\usepackage{lscape}
\usepackage{url}
\usepackage{multirow}
\usepackage{array}
\usepackage{paralist}
\usepackage{makecell}
\usepackage{adjustbox}

\usepackage{braket}
\usepackage[framemethod=tikz]{mdframed}
\usepackage{subcaption}

\mdfdefinestyle{mpdframe}{
    frametitlebackgroundcolor   =black!15,
    frametitlerule              =true,
    roundcorner                 =5pt,
    middlelinewidth             =0.8pt,
    innermargin                 =0.2cm,
    outermargin                 =0.2cm,
    innerleftmargin             =0.2cm,
    innerrightmargin            =0.2cm,
    innertopmargin              =0.2cm,
    innerbottommargin           =0.2cm
}

\usepackage{xcolor}

\usepackage{xurl}
\usepackage{algorithmic}

\usepackage{float}
\floatstyle{plaintop}
\restylefloat{table}

\usepackage{array}
\usepackage{stfloats}
\usepackage{url}
\usepackage{tabularx}
\usepackage{xspace}
\usepackage{graphicx}
\usepackage{epstopdf} 
\usepackage{wrapfig}
\usepackage{pbox}

\usepackage{tikz}
\usepackage{soul}
\usetikzlibrary{arrows.meta}
\usetikzlibrary{arrows}
\usetikzlibrary{shapes}
\usetikzlibrary{patterns} 
\usetikzlibrary{positioning}
\usetikzlibrary{calc}
\usepackage{ragged2e}

\usepackage{amsmath}
\newtheorem{metric}{Metric}

\usepackage[colorinlistoftodos,prependcaption]{todonotes}
\newboolean{showcomments}
\setboolean{showcomments}{true}
\ifthenelse{\boolean{showcomments}}
 { \newcommand{\mynote}[2]{
      \fbox{\bfseries\sffamily\scriptsize#1}
        {\small$\blacktriangleright$\textsf{\textcolor{red}{{\em #2}\bf }}$\blacktriangleleft$}}}
        { \newcommand{\mynote}[2]{}}

\usepackage{colortbl}
\definecolor{pblue}{rgb}{0.13,0.13,1}
\definecolor{pgreen}{rgb}{0,0.5,0}
\definecolor{pred}{rgb}{0.9,0,0}
\definecolor{pgrey}{rgb}{0.46,0.45,0.48}

\usepackage{listings}
\usepackage[scaled]{beramono}
\usepackage{tikz}
\usepackage{xspace}
\usepackage{listings}
\lstdefinelanguage{Pom}{
  morekeywords={<dependency>,</dependency>,<groupId>,</groupId>,<artifactId>,</artifactId>,<version>,</version>,<scope>,</scope>},
  otherkeywords={<dependency>,</dependency>,<groupId>,</groupId>,<artifactId>,</artifactId>,<version>,</version>,<scope>,</scope>}
}

\usetikzlibrary{calc}
\usetikzlibrary{decorations.pathmorphing}

\usepackage[colorinlistoftodos,prependcaption]{todonotes}
\usepackage{booktabs}
\def\BibTeX{{\rm B\kern-.05em{\sc i\kern-.025em b}\kern-.08emT\kern-.1667em\lower.7ex\hbox{E}\kern-.125emX}}

\makeatletter
\pgfarrowsdeclare{crow's foot}{crow's foot}
{
    \pgfarrowsleftextend{+-.5\pgflinewidth}%
    \pgfarrowsrightextend{+.5\pgflinewidth}%
}
{
    \pgfutil@tempdima=0.6pt%
    \pgfsetdash{}{+0pt}%
    \pgfsetmiterjoin%
    \pgfpathmoveto{\pgfqpoint{0pt}{-9\pgfutil@tempdima}}%
    \pgfpathlineto{\pgfqpoint{-13\pgfutil@tempdima}{0pt}}%
    \pgfpathlineto{\pgfqpoint{0pt}{9\pgfutil@tempdima}}%
    \pgfpathmoveto{\pgfqpoint{0\pgfutil@tempdima}{0\pgfutil@tempdima}}%
    \pgfpathmoveto{\pgfqpoint{-8pt}{-6pt}}%
    \pgfpathlineto{\pgfqpoint{-8pt}{-6pt}}%
    \pgfpathlineto{\pgfqpoint{-8pt}{6pt}}%
    \pgfusepathqstroke%
}

\pgfarrowsdeclare{omany}{omany}
{
    \pgfarrowsleftextend{+-.5\pgflinewidth}%
    \pgfarrowsrightextend{+.5\pgflinewidth}%
}
{
    \pgfutil@tempdima=0.6pt%
    \pgfsetdash{}{+0pt}%
    \pgfsetmiterjoin%
    \pgfpathmoveto{\pgfqpoint{0pt}{-9\pgfutil@tempdima}}%
    \pgfpathlineto{\pgfqpoint{-13\pgfutil@tempdima}{0pt}}%
    \pgfpathlineto{\pgfqpoint{0pt}{9\pgfutil@tempdima}}%
    \pgfpathmoveto{\pgfqpoint{0\pgfutil@tempdima}{0\pgfutil@tempdima}}%
    \pgfpathmoveto{\pgfqpoint{0\pgfutil@tempdima}{0\pgfutil@tempdima}}%
    \pgfpathmoveto{\pgfqpoint{-6pt}{-6pt}}%
    \pgfusepathqstroke%
    \pgfsetfillcolor{white}
    \pgfpathcircle{\pgfpoint{-11.5pt}{0}} {3.5pt}
    \pgfusepathqfillstroke%
}

\pgfarrowsdeclare{one}{one}
{
    \pgfarrowsleftextend{+-.5\pgflinewidth}%
    \pgfarrowsrightextend{+.5\pgflinewidth}%
}
{
    \pgfutil@tempdima=0.6pt%
    \pgfsetdash{}{+0pt}%
    \pgfsetmiterjoin%
    \pgfpathmoveto{\pgfqpoint{0\pgfutil@tempdima}{0\pgfutil@tempdima}}%
    \pgfpathmoveto{\pgfqpoint{-6pt}{-6pt}}%
    \pgfpathlineto{\pgfqpoint{-6pt}{-6pt}}%
    \pgfpathlineto{\pgfqpoint{-6pt}{6pt}}%
    \pgfpathmoveto{\pgfqpoint{0\pgfutil@tempdima}{0\pgfutil@tempdima}}%
    \pgfpathmoveto{\pgfqpoint{-8pt}{-6pt}}%
    \pgfpathlineto{\pgfqpoint{-8pt}{-6pt}}%
    \pgfpathlineto{\pgfqpoint{-8pt}{6pt}}%
    \pgfusepathqstroke%
}

\pgfarrowsdeclare{oone}{oone}
{
    \pgfarrowsleftextend{+-.5\pgflinewidth}%
    \pgfarrowsrightextend{+.5\pgflinewidth}%
}
{
    \pgfutil@tempdima=0.6pt%
    \pgfsetdash{}{+0pt}%
    \pgfsetmiterjoin%
     \pgfpathmoveto{\pgfqpoint{0\pgfutil@tempdima}{0\pgfutil@tempdima}}%
    \
    pgfpathmoveto{\pgfqpoint{-4pt}{-6pt}}%
    \pgfpathlineto{\pgfqpoint{-4pt}{-6pt}}%
    \pgfpathlineto{\pgfqpoint{-4pt}{6pt}}%
    \pgfsetfillcolor{white}
    \pgfpathcircle{\pgfpoint{-11.5pt}{0}} {3.5pt}
    \pgfusepathqfillstroke%
}
\makeatother

\tikzset{%
    mylabel/.style={font=\footnotesize},
    pics/entity/.style n args={3}{code={%
        \node[draw,
        rectangle split,
        rectangle split parts=2,
        text height=1.5ex,
        text width=7em,
        text centered
        ] (#1)
        {#2 \nodepart[font=\small, text centered]{second}
            \begin{tabular}{>{\raggedright\arraybackslash}p{9em}}
                #3
            \end{tabular}
        };%
    }},
    zig zag to/.style={
        to path={(\tikztostart) -| ($(\tikztostart)!#1!(\tikztotarget)$) |- (\tikztotarget)}
    },
    zig zag to/.default=0.5,   
    one to one/.style={
        one-one, zig zag to
    },
    one to oone/.style={
        one-oone, zig zag to
    },
    oone to none/.style={
        oone-, zig zag to
    },
    oone to oone/.style={
        oone-oone, zig zag to
    },
    one to many/.style={
        one-crow's foot, zig zag to,
    },
    one to omany/.style={
        one-omany, zig zag to
    },
    one to none/.style={
        one-, zig zag to
    },   
    many to many/.style={
        crow's foot-crow's foot, zig zag to
    }
}

\usetikzlibrary{calc}
\usetikzlibrary{decorations.pathmorphing}

\makeatletter

\newcommand{\defhighlighter}[3][]{%
  \tikzset{every highlighter/.style={color=#2, fill opacity=#3, #1}}%
}

\defhighlighter{yellow}{.5}

\newcommand{\highlight@DoHighlight}{
  \fill [ decoration = {random steps, amplitude=1pt, segment length=15pt}
        , outer sep = -15pt, inner sep = 0pt, decorate
        , every highlighter, this highlighter ]
        ($(begin highlight)+(0,8pt)$) rectangle ($(end highlight)+(0,-3pt)$) ;
}

\newcommand{\highlight@BeginHighlight}{
  \coordinate (begin highlight) at (0,0) ;
}

\newcommand{\highlight@EndHighlight}{
  \coordinate (end highlight) at (0,0) ;
}

\newdimen\highlight@previous
\newdimen\highlight@current

\DeclareRobustCommand*\highlight[1][]{%
  \tikzset{this highlighter/.style={#1}}%
  \SOUL@setup
  \def\SOUL@preamble{%
    \begin{tikzpicture}[overlay, remember picture]
      \highlight@BeginHighlight
      \highlight@EndHighlight
    \end{tikzpicture}%
  }%
  \def\SOUL@postamble{%
    \begin{tikzpicture}[overlay, remember picture]
      \highlight@EndHighlight
      \highlight@DoHighlight
    \end{tikzpicture}%
  }%
  \def\SOUL@everyhyphen{%
    \discretionary{%
      \SOUL@setkern\SOUL@hyphkern
      \SOUL@sethyphenchar
      \tikz[overlay, remember picture] \highlight@EndHighlight ;%
    }{%
    }{%
      \SOUL@setkern\SOUL@charkern
    }%
  }%
  \def\SOUL@everyexhyphen##1{%
    \SOUL@setkern\SOUL@hyphkern
    \hbox{##1}%
    \discretionary{%
      \tikz[overlay, remember picture] \highlight@EndHighlight ;%
    }{%
    }{%
      \SOUL@setkern\SOUL@charkern
    }%
  }%
  \def\SOUL@everysyllable{%
    \begin{tikzpicture}[overlay, remember picture]
      \path let \p0 = (begin highlight), \p1 = (0,0) in \pgfextra
        \global\highlight@previous=\y0
        \global\highlight@current =\y1
      \endpgfextra (0,0) ;
      \ifdim\highlight@current < \highlight@previous
        \highlight@DoHighlight
        \highlight@BeginHighlight
      \fi
    \end{tikzpicture}%
    \the\SOUL@syllable
    \tikz[overlay, remember picture] \highlight@EndHighlight ;%
  }%
  \SOUL@
}

\makeatother
\newcommand{\ie}{i.e.\@\xspace}

\newcommand{\etc}{etc.\@\xspace}

\newcommand{\pom}{\textit{pom.xml}\@\xspace}

\newcommand{\mc}{Maven Central\@\xspace}
\newcommand{\mdg}{MDG\@\xspace}

\newcommand{\lib}{\texttt{LIB}\@\xspace}
\newcommand{\libs}{\texttt{LIBs}\@\xspace}

\newcommand{\mlib}{\text{\lib}}

\newcommand{\flink}{\textit{org\-.apache\-.flink\-:flink\-core\-:1.5.1}\@\xspace}
\newcommand{\flinkRun}{\textit{org\-.apache\-.flink\-:flink\-runtime\-:1.5.0}\@\xspace}

\newcommand{\jsr}{\textit{com\-.google\-.code\-.findbugs\-:jsr305\-:1.3.9}\@\xspace}
\newcommand{\jsrga}{\textit{jsr305\-:1.3.9}\@\xspace}
\newcommand{\flinkClass}{\textit{org\-.apache\-.flink\-.runtime\-.entrypoint\-.Cluster\-Entry\-point}\@\xspace}

\newcommand{\fuserst}{clients_{obs}(type)}
\newcommand{\ftypes}{types(library)}
\newcommand{\ftypeso}{types_{obs}(library)}
\newcommand{\fusers}{clients(library)}
\newcommand{\fuserso}{clients_{obs}(library)}

\newcommand{\fstr}{TUR(type)}

\newcommand{\jar}{\textit{jar}\@\xspace}

\newcommand{\slfgav}{\textit{org\-.slf4j\-:slf4j-api\-:1.7.21}\@\xspace}

\newcommand{\commons}{\textit{com\-mons-cli\-:1.3.1}\@\xspace} 
\newcommand{\javax}{\textit{javax\-.inject\-:1}\@\xspace}
\newcommand{\junit}{\textit{junit:4.12}\@\xspace}
\newcommand{\hibernate}{\textit{hiber\-nate-core\-:4.3.11\-.Final}\@\xspace}

\newcommand{\slf}{\textit{slf4j-api\-:1.7.21}\@\xspace}
\newcommand{\comio}{\textit{commons-io:2.4}\@\xspace}

\newcommand{\nblibraryga}{94\@\xspace}
\newcommand{\nblibrarygav}{5,225\@\xspace}
\newcommand{\nbclientga}{99,949\@\xspace}
\newcommand{\nbrclientgav}{829,410\@\xspace}
\newcommand{\nbclientgav}{901,876\@\xspace} 
\newcommand{\nbdependency}{2,376,526\@\xspace}

\newcommand{\nbrdependency}{2,169,273\@\xspace} 
\newcommand{\nbrdependencynotobs}{892,167\@\xspace}

\newcommand{\nbrdependencyobs}{1,277,106\@\xspace}
\newcommand{\nblibrarygavobs}{4,931\@\xspace}
\newcommand{\nbrclientgavobs}{677,953\@\xspace}

\newcommand{\nbmostpopcli}{235,440\@\xspace}
\newcommand{\nbmostpopdep}{319,882\@\xspace}

\newcommand{\nbapielementsused}{5,076,307\@\xspace}

\newcommand{\nbusage}{87,207,807\@\xspace}
\newcommand{\percdepused}{58.87\%\@\xspace}
\newcommand{\percdepnotused}{41,13\%\@\xspace}

\usepackage[numbers]{natbib}

\newcommand{\modified}[1]{{#1}}

\def\tsc#1{\csdef{#1}{\textsc{\lowercase{#1}}\xspace}}
\tsc{WGM}
\tsc{QE}
\tsc{EP}
\tsc{PMS}
\tsc{BEC}
\tsc{DE}

\begin{document}
\let\WriteBookmarks\relax
\def\floatpagepagefraction{1}
\def\textpagefraction{.001}

\shorttitle{API Beauty is in the eye of the Clients}

\title [mode = title]{API Beauty is in the eye of the Clients: 2.2 Million Maven Dependencies reveal the Spectrum of Client-API Usages}                      

\shortauthors{Harrand et~al.}

\author[1]{Nicolas Harrand}[orcid=0000-0002-2491-2771]\ead{harrand@kth.se}
\cormark[1]

\author[2]{Amine Benelallam}[orcid=0000-0003-3064-8302]\ead{amine.benelallam@inria.fr}

\author[1]{C\'esar Soto-Valero}[orcid=0000-0003-0541-6411]\ead{cesarsv@kth.se}

\author[3]{Fran\c{c}ois Bettega}[orcid=0000-0002-9736-5289]\ead{francois.bettega@univ-grenoble-alpes.fr}

\author[2]{Olivier Barais}[orcid=0000-0002-4551-8562]\ead{barais@irisa.fr}

\author[1]{Benoit Baudry}[orcid=0000-0002-4015-4640]\ead{baudry@kth.se}

\cortext[cor1]{Corresponding author}

\address[1]{KTH Royal Institute of Technology, Stockholm, Sweden}
\address[2]{Univ Rennes, Inria, CNRS, IRISA, Rennes, France}
\address[3]{Univ. Grenoble Alpes, Inserm, CHU Grenoble Alpes, HP2, Grenoble, France}


\begin{abstract}
Hyrum's law states a common observation in the software industry: ``With a sufficient number of users of an API, it does not matter what you promise in the contract: all observable behaviors of your system will be depended on by somebody''. Meanwhile, recent research results seem to contradict this observation when they state that ``for most APIs there is a small number of features that are actually used''. 
We investigate this seeming paradox between the observations in industry and the research literature, with a large scale empirical study of client API relationships in one single ecosystem: Maven central. 
    
\modified{We study the \nblibraryga most popular libraries in \mc, as well as the \nbrclientgav client artifacts that declare a dependency to these libraries and that are available in \mc, summing up to 2.2M dependencies.}
Our analysis indicates the existence of a wide spectrum of  API usages, with enough clients most API types end up being used at least once.
Our second key observation is that, for all libraries, there is a small set of API types that are used by the vast majority of its clients. 
\modified{The practical consequences of this study are two-fold: (i) it is possible for API maintainers to find an essential part of their API on which they can focus their efforts; (ii) API developers should limit the public API elements to the set of features for which they are ready to have users.}
\end{abstract}




\maketitle

\printcredits

\sloppy

\section{Introduction}  \label{sec:introduction}

Software libraries  provide reusable functionalities via Application Programming Interfaces (APIs). Software applications that reuse these functions in their code declare the list of APIs they wish to use. This declaration creates a \emph{dependency} between the \emph{client} application and the \emph{library} API.
Our study  focuses on two well-documented intuitions about such client-library dependencies.
On one hand, Hyrum's law captures a common observation in the software industry: ``\emph{With a sufficient number of users of an API, ...all observable behaviors of your system
will be depended on by somebody}'' \cite{Hyrums2019}. Applied to API usages, this would suggest that with enough clients, even the most exotic API elements would eventually be used by at least a client.
On the other hand, recent research results concur to consolidate the intuition that APIs are unnecessarily large and that client dependencies actually focus on a small part of common APIs \cite{qiu2016understanding,Eghan2019,Sawant2017,Mastrangelo2015}.

We are intrigued by the seeming contradiction between these observations: the first one suggests that every API member is eventually used, while the second one suggests that only a small part of APIs is really necessary. Our analysis of millions of dependencies reveals a continuum between these two extremes rather than a contradiction. In other words, we confirm that libraries contain a portion of API types that are used by a vast majority of clients. Meanwhile, in the presence of a sufficient number of clients, we observe that the rare or exotic API types would eventually fit at least one adventurous client.
\modified{
These results provide evidence that API developers can make trade-offs between the share of API elements they consider in maintenance, documentation, and automated migration tools and the share of clients that they support.  To support all clients, developers need to invest effort that is proportional to the total size of APIs. Yet, accepting to support only the majority of clients, which use the core API types, allows for significant effort savings. Considering the typical API maintenance task of migration \cite{bartolomei2009}, the migration of the $6317$ clients of library \texttt{gson:2.3.1} to \texttt{jackson-databind} requires supporting migration rules for $162$ API types, but this number of types can be decreased to $20$ ($12\%$) if only $5669$ ($90\%$) clients are to be supported.}


Our work explores this spectrum of dependency relations, focusing on the \mc ecosystem. This choice is motivated by two factors: it is the most popular repository to distribute code artifacts that run on the Java Virtual Machine;  it contains both APIs and clients that depend on these APIs. The Maven Dependency Graph \cite{Benelallam2019} provides a snapshot of Maven Central as of September 6, 2018. From this graph, we determine the \nblibraryga most used libraries and \modified{all the client artifacts in the repository,} that depend on any version of one of these libraries. This forms the dataset for our study: \nblibrarygav libraries (union of all versions of the \nblibraryga most popular libraries), \nbclientgav clients, summing up to  \nbrdependency dependencies.

\modified{We study how Maven artifacts depend on each other, around three dimensions.} First, we analyze the client-side, to determine to what extent each declared dependency is actually used at least once in their code, i.e., there is at least one API member used by the client's code. 
Second, we analyze the API side of dependencies to determine how different API types are used by the clients. We split this analysis into two steps: we start by investigating the usages of an API, cumulating all its versions; later, we analyze the most popular version of each API.
\modified{Finally, to put our findings to use, we propose a new actionable way to explore the continuum of dependencies and assess the impact of focusing on a small subset of APIs, e.g., for maintenance activities through extinction sequences. For tasks where costs and effort increase with the size of APIs such as API migration \cite{bartolomei2009}, a trade\-off can be made between cost and number of clients supported.}

The key findings of our study are as follows: 
(i) on the clients-side, we found that \percdepnotused of declared dependencies do not translate into API usages at the bytecode-level;
(ii) on the libraries-side, we observe the following: when considering the most used version of each library, it is very likely that every public member is used; 
(iii) meanwhile, we notice that every API can be reduced to a small fraction and still fulfill the needs of a majority of the clients.
The size of this fraction varies from one API to another, as library API purpose, size, and usage differ.
Our dataset is large enough to include some of the most extreme cases that occur in the extraordinary practice of software development, e.g., a very small API with only annotations, some giant APIs which clients use in a very focused way, or even some artifacts that are massively used even if they have no public types.

The contributions of this paper are as follows:
\begin{itemize} 
    \item A systematic large-scale analysis of  of \nbrdependency Maven client - API relations.
    \item \modified{A public dataset of \nblibrarygav libraries (union of all versions of the \nblibraryga most popular libraries) and  \nbclientgav clients drawn from Maven Central~\cite{Zenodo2019} along with an open reproduction package~\cite{CodeRepo2019}.} This large dataset can fuel the ongoing research initiatives in the areas of dependency management and release engineering. 
    \item Novel empirical evidence about Maven dependencies: all APIs include a small set of types that is used by the majority of their clients, while most API types are used by at least one client, when considering the most popular version of an API. \modified{These findings open new directions to improve Maven's build process and to focus effort on the relevant subset of APIs for  maintenance and migration tasks.}
\end{itemize}

This paper is organized as follows. Section \ref{sec:background} introduces the key concepts of Maven. In Section~\ref{sec:methodology} we present our research methodology, analysis infrastructure and the dataset for this study. In Section~\ref{sec:results} we discuss the empirical observations about the actual usage of client-library Maven dependencies. In Section~\ref{sec:discussion} we discuss how our results could generalize in other ecosystems.

\section{API usage in the Maven ecosystem}
\label{sec:background}

Maven is a software project management tool for Java and other languages targeting the JVM (e.g., Groovy, Kotlin, Clojure, Scala). It automates most phases of a software development life-cycle, from build to deployment. Maven relies on a specification file, named \pom, where developers explicitly declare what should happen at each building phase. Dependency management is one important phase where Maven automatically fetches software artifacts on which a project depends. Those artifacts are hosted on remote repositories, either public or private. Currently, \mc is the most popular public repository. It hosts millions of software artifacts coming in the form of binary sources (e.g., \jar). These artifacts are uniquely identified by GAV coordinates, referring to groupId (G), artifactId (A), and version (V). Artifacts in \mc cannot be modified or updated, meaning that all the releases of each artifact are stored in the repository. \looseness=-1

Maven lets developers the possibility to specify the scope of a dependency declaration. This scope determines when the \jar of the dependency is added to the classpath of a project~\cite{MavenScope}.
\textit{Compile} is the default scope, which implies that the dependency is added to the compilation, test, and runtime classpaths. A dependency with \textit{Compile} scope is also added transitively to the classpath of the artifact's clients (and their clients and so forth), if they do not explicitly exclude it.
The dependencies declared with a \textit{Test} are meant to be required only to test the project: they are added only to the compilation and test classpaths; they are not added as transitive dependencies for the clients. 
The  scopes \textit{Provided} and \textit{System} indicate that the dependency will be provided directly by the user if needed at runtime. Such dependencies are resolved by Maven only for the compilation and test classpath, and are not transitive. The scope \textit{Provided} can be used for a dependency that is required only to build a project but not necessary for its execution.
Finally, a \textit{Runtime} dependency is only added to the runtime classpath and is transitive.

\begin{figure}[h]
    \centering
    \includegraphics[origin=c,width=1\columnwidth]{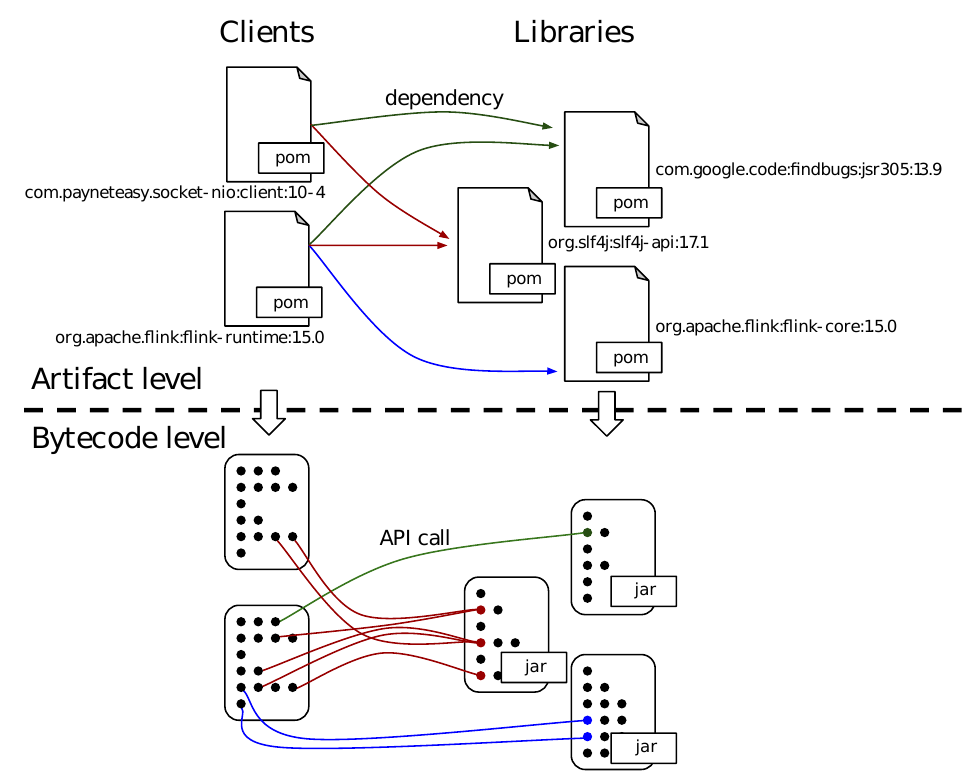}
    \caption{Software reuse principles in JVM-based projects.}
    \label{fig:software-reuse}
\end{figure}

\autoref{fig:software-reuse} illustrates a simplified example of API usages within the Maven ecosystem. API usages happen at two levels, the artifact-level, and the code-level. At the artifact level, a project declares a list of libraries that have to be added to the project's classpath in order to build correctly. At the code-level, the members of the  API (e.g., types, methods, etc.) are called, e.g., via object instantiation.

\subsection{Artifact-level API dependency}

\autoref{fig:software-reuse} represents the dependency relationships between five artifacts. The \textit{com.\-payneteasy.\-socket-nio:\-client:\-1.0-4} artifact declares dependencies towards two libraries: \textit{com.\-google.\-code:\-findbugs:jsr305:\-13.9} and \textit{org.slf4j:\-slf4j-api:\-1.7.1}. The \textit{org.\-apache.\-flink:\-flink-runtime:\-1.5.0} artifact declares a dependency towards three libraries: \textit{com.\-google.\-code:\-findbugs:\-jsr305:\-13.9}, \textit{org.\-slf4j:\-slf4j-api:\-1.7.1} and \textit{org.\-apache.\-flink:\-flink-core:\-1.5.0}.\sloppy

The dependencies declared in the \pom file of each artifact are identified by their exact coordinates. For example, in \autoref{lst:flink-pom}, the artifact \flinkRun declares a dependency towards \jsr ~to reuse the javax annotations defined in this library. Consequently, when building the \textit{flink-runtime} project, Maven will fetch the resource \jar corresponding to \jsrga, together with all its transitive dependencies, and add them to the project's classpath. 

\begin{lstlisting}[basicstyle=\footnotesize\ttfamily,language=Pom, float, caption={Excerpt of the \pom file of flink-runtime:1.5.0}, label={lst:flink-pom}, numbers=left, linewidth=0.92\columnwidth]  
<dependency>
    <groupId>com.google.code.findbugs</groupId>
    <artifactId>jsr305</artifactId>
    <version>1.3.9</version>
    <scope>compile</scope>
</dependency>
\end{lstlisting}

\subsection{Code-level API dependency}

\input{figures/code.tex}

There are many ways to use external APIs at the code level through inheritance, implementation, composition, genericity, static method invocation \etc.  \autoref{lst:client-ex} shows a snippet from the class \flinkClass of the library \flinkRun~\cite{Flink2019}. It illustrates different ways in which \flinkRun uses some dependencies that are declared in its \pom. The class \textit{Cluster\-Entrypoint} implements the \textit{Auto\-Close\-able\-Async} class exposed by the \flink dependency (line~\ref{line:implements}), while lines~\ref{line:logger}~and~\ref{line:encaps} are examples of field declarations. On line~\ref{line:logger} the dependency is used through a call to the static method \textit{getLogger()}. Lines~\ref{line:rem-start}-\ref{line:rem-end} illustrate  reuse examples of API members such as annotations, methods call, or in methods signatures.


\section{Methodology}
\label{sec:methodology}


In this section we present our research questions and the metrics we use to answer these questions. Then, we introduce the dataset for this study and the methodology we have used to collect API usages.

\subsection{Research questions}


This study is structured around the following research questions:

\newcommand{\RQone}{\modified{How are the APIs used in the code of clients that declare a dependency towards them?}\xspace}


\newcommand{\RQtwo}{How is the usage frequency of API types distributed and how does that depend on the number of clients?\xspace}

\newcommand{\RQthree}{How is the usage frequency of API types distributed when focusing on the popular version of an API?\xspace} 

\newcommand{\RQfour}{Can inter-package calls explain the existence of API types that are unused by the clients?\xspace}

\newcommand{\RQfive}{How many API classes are essential for most of the clients?\xspace}

\textbf{RQ1: \RQone} 
In this work we study how the APIs are used in the code of clients that declare a dependency towards them. Previous studies hint on the fact that some declared dependencies are actually never used, for example, because API users do not systematically maintain their build file \cite{Constantinou2014,Zaimi2015}.
In this research question we investigate to what extent there is a gap between what clients declare in their \pom and what they actually use in their code. We also investigate what causes these discrepancies.
\newpage

\textbf{RQ2: \RQtwo}  API developers aim at providing reusable functionalities to a large number of clients. This desire to satisfy many users can become a double-edged sword from the users' perspective, which can be overwhelmingly loaded by a large number of API types that they do not need~\cite{Piccioni2013,Myers2016}. 
This research question focuses on the distribution of usage frequency of API types. We measure the share of API types that are never used by any clients, rarely used or used by most clients. We discuss how these shares vary depending on the number of clients. For this question we only consider Library \- Client relationship were the client uses at least one type of the API.

\textbf{RQ3: \RQthree}
In this question we focus on the most popular version of each library in our dataset to determine the effect of a ``sufficient number of users'' on API usage. This sheds new light on usages ratios, compared to RQ2 that considers all versions of the libraries. This question is at the core of our analysis of API usage with respect to Hyrum's law,  which requires a large number of clients to study.

\textbf{RQ4: \RQfour} 
In Java, developers need to set the visibility of classes or methods to  public if they want to allow inter-package usages. In other words, some parts of a library's API might be public only because they are intended to be used by other package of the library, and might not be meant to be reused by the library's clients. 
In this research question, we analyze whether inter-package usages indeed contribute to explain the existence of  API types that are not used by the clients.

\textbf{RQ5: \RQfive} The usage of API types is demonstrated to be strongly related to the needs of the clients~\cite{Sawant2018}. In the long term, these needs determine what constitutes the essential part of an API. Here, we address the key intuition of this work: the existence of a \emph{reuse-core} for the APIs, i.e. a set of highly used elements according to the clients' state of practice. In this research question, we investigate what proportion of the API is essential for the clients and how this reuse-core varies according to various API usages.

\subsection{Metrics and Definitions}
\label{sec:metrics}

For further references, we introduce the following notations: 
\begin{itemize}
    \item $library$: an artifact declared as a dependency by a $client$
    \item $\ftypes$: the set of distinct types that are visible for $client$ elements, \ie classes, interfaces, or annotations \item $\ftypeso$ is the subset of $\ftypes$ used by at least one client    
    \item \lib a set of libraries sharing the same groupId and artifactId, regardless of the version.
    \item $\fusers$: the set of clients that declare a dependency towards a $library$
    \item $\fuserso$ is the set of $client$ that call at least one element of $\ftypeso$
    \item $\fuserst$ is the set of $clients$ that call at least one member, \ie fields and public and protected methods, including constructors, of a given $type$ 
\end{itemize}

To answer RQ1, we measure the possible gap between clients that declare a dependency towards an API in their \pom and the ones that actually call this API at least once in their bytecode.

\begin{metric}\label{metric:dur}
The dependency usage rate (DUR) of a  \lib is the proportion of clients that call at least one API member of a $library \in \mlib$, (observed through static analysis), among all the clients that declare a dependency towards any version of \lib:
\begin{equation*}
     DUR(\mlib) = \frac{\mid\displaystyle\bigcup_{l \in \mlib}clients_{obs}(l)\mid}{\mid\displaystyle\bigcup_{l \in \mlib}clients(l)\mid}
\end{equation*}

\end{metric}

RQs 2, 3 and 4 study what proportion of the clients of a libraries use each type of its API. We consider that a client uses a type if it uses at least one member of this type, i.e. $client \in \fuserst$. We name this proportion type usage rate (TUR), and define it as follows.

\begin{metric}\label{metric:str}
The type usage rate (TUR) of a given $type \in library$ corresponds to the proportion of clients that reference at least one member of said $type$ (observed through static analysis), \ie $\fuserst$, among the clients that actually use $library$, \ie $\fuserso$: 
\begin{equation*}
    \fstr = \frac{\mid\fuserst\mid}{\mid\fuserso\mid}
\end{equation*}
\end{metric}

RQ5 investigates how necessary is each type of the API of the most popular version of each \lib.  To assess this necessity, we adapt the concept of extinction sequence \cite{albert2000error} to simulate the hiding of each type $\in \ftypes$ from the least used to the most used. We call $LU(library,n)$ the set of $n\%$ least used types in $\ftypeso$.

\begin{metric} \label{metric:ext-seq} 
We measure the surviving client share (SCS) unaffected by the hiding of $LU(library,n)$.
\begin{equation*}
SCS(library,n) = \frac{
\big|
\Set{ c |  \begin{array}{l} \forall type \in LU(library,n),\\
c \notin \fuserst\end{array}}
\big|
}{\mid\fuserso\mid}
\end{equation*}
\end{metric}

To answer RQ5, we compute the variation of $SCS(library, n)$, where we vary $n$ from $0\%$ to $100\%$.

\subsection{Dataset}
\label{sec:data}
In this work, we analyze software dependencies both at the artifact and at the code levels. At the artifact level, we analyze \pom files of client projects to determine the list of direct dependencies they declare. At the code level, we analyze the bytecode of both the clients and the libraries. On the client-side, we determine what parts of the libraries' API they  actually use. On the library-side, we evaluate the extent to which an API is actually used by its clients. Hence our dataset is composed of bytecode and pom file of both libraries and clients.

We leverage the Maven Dependency Graph (MDG)~\cite{Benelallam2019} to identify the most popular APIs  in \mc, \modified{as well as their client artifacts.} Then we extract usage information through static analysis of the \jar artifacts. This section details these two steps.

The \mdg captures all artifacts in \mc as nodes and their dependencies as directed edges. Every node has a \textit{coordinates} property referring to the artifact's coordinates (GAV) and a packaging referring to the format of the artifacts binaries. Furthermore, every edge has a property \textit{scope} identifying the dependency scope. We extract the $100$ most popular libraries. 
We exclude $6$ \libs from these $100$ libraries. One of them, \texttt{appcompat-v7}, is not packaged as a jar but as an apk. The $5$ other ignored \libs contain no type and so, no API usage can be observed in their clients. Three of these $5$ libraries are written in Clojure, and two others, \texttt{spring-\allowbreak{}boot-\allowbreak{}starter} and \texttt{spring-\allowbreak{}boot-\allowbreak{}starter-\allowbreak{}web}, are packaged as jar files that contain no bytecode, i.e., there are no API types. In fact, these \libs serve as an alias for a group of commonly used dependencies that are transitively inherited through a single entry-point. 
Hence, we study a set of $94$ libraries. We compute the popularity of a \lib based on the number of distinct clients that declare a dependency toward a version of the \lib with a \textit{Compile} scope.


The raw dataset for our study includes all dependency relationships from any \modified{$client$ artifact, in Maven Central,} towards any version of one of our \libs. This represents \nbdependency dependency relationship between \nbclientgav clients (belonging to \nbclientga unique pairs \textit{(GroupID, ArtifactID)}) and \nblibraryga \libs. The \libs are in a total of \nblibrarygav versions in the dataset.

\begin{figure}[t]
  \centering
  \includegraphics[width=01\columnwidth]{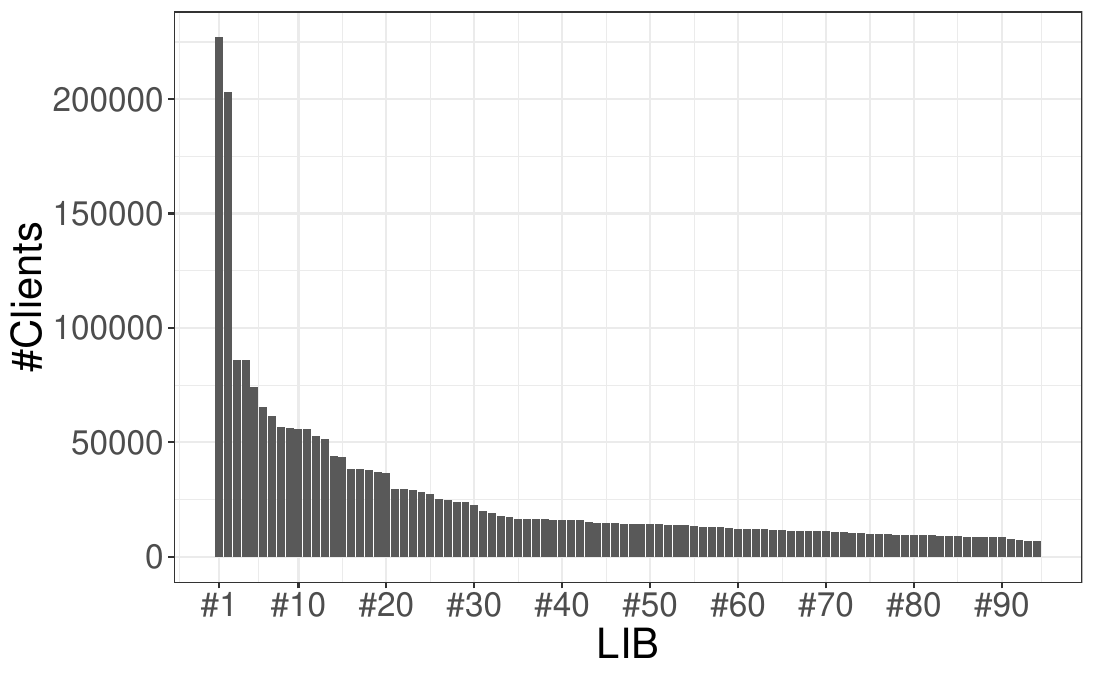}
  \caption{Distribution of LIB number of clients}
  \label{fig:lib-popularity}
\end{figure}

\autoref{fig:lib-popularity} shows the distribution of the number of clients for each LIB. The two most popular libraries are the standard \texttt{scala-library} and \texttt{slf4j-api} with respectively 227, 014 and 203,366 clients. The number of clients per library decreases quickly in this ranking to reach 7,007 clients for the least popular library of our dataset, \texttt{xercesImpl}.
These libraries cover a broad range of application domains, from logging, networking, language extensions, to collections and bytecode manipulation.
Our dataset includes libraries from $15$ of the $20$ most popular categories of libraries from \textit{mvnrepository.com}\footnote{\url{https://mvnrepository.com/open-source}}. The only categories not covered are, two related to Android applications (since we exclude apk), two related to testing (as we exclude \textit{test} dependencies), and a category related to web assets which do not contain bytecode.

\begin{figure}[t]
  \centering
  \includegraphics[width=01\columnwidth]{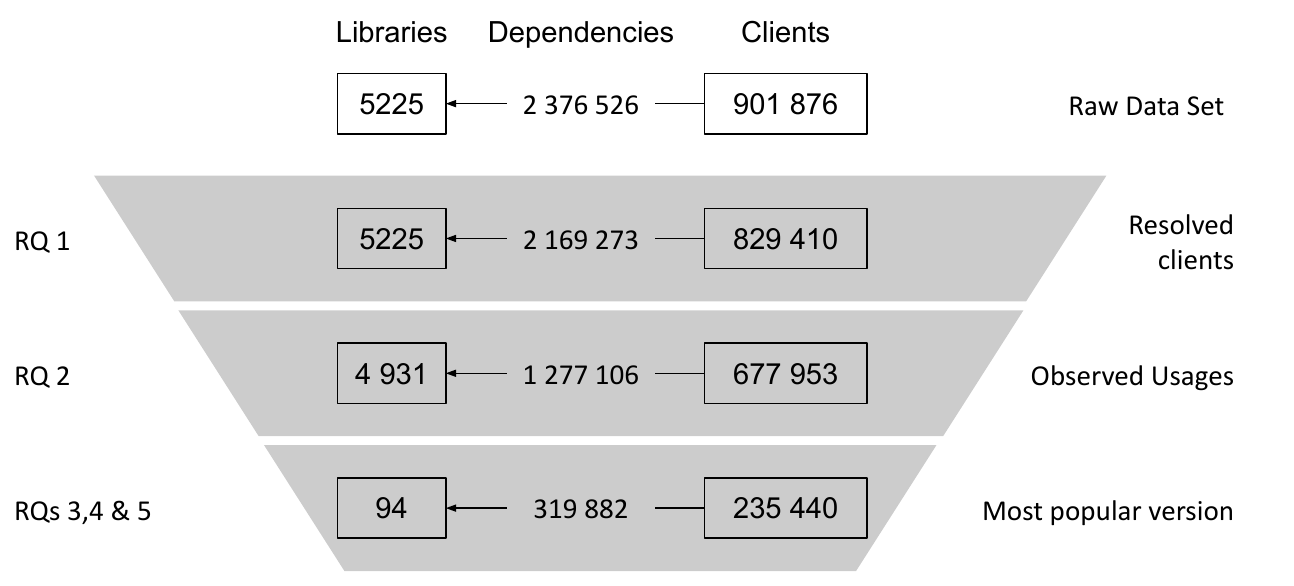}
  \caption{Progressive data set filtering.}
  \label{fig:data-filtering}
\end{figure}

As illustrated in \autoref{fig:data-filtering}, we filter our dataset through the research questions. For RQ1, we focus on the \nbrdependency dependencies concerning  the \nbrclientgav client artifacts that we could resolve (those for which we could download the jar). For RQ2 we focus on the dependencies for which we could observe an actual usage in the bytecode of the client. At this stage we exclude 2 LIB that do not contain public types. This represents \nblibrarygavobs libraries, \nbrdependencyobs dependencies and \nbrclientgavobs clients. In RQs 3, 4 and 5 we analyze the client - API dependencies for the  most popular version of each library. This corresponds to $94$ libraries, \nbmostpopdep dependency relationship and \nbmostpopcli unique clients. This latest version of the dataset supports our investigations of API usage with ``a sufficient number of users'', a key condition to study long tail distributions.

Given the large number of libraries and clients, the plots displayed in the section represent a lot of information and it is sometimes difficult to keep the intuition between the data and the software engineering phenomena that are at stake. To keep the discussion concrete, we select 6 libraries that we use to illustrate all the research questions. \autoref{tab:six-libs} summarizes the name, the number of types, the number of clients, the number of clients that actually use the library and the application domain for these 6 libraries. We select these libraries because they represent a diverse set of domains, sizes,  API types, and number of clients. We select the most used version of each \lib. 
\looseness=-1

\begin{table}[ht]
    \scriptsize
    \setlength{\tabcolsep}{4.5pt}
    \centering
    \begin{tabularx}{\columnwidth}{lllll}
      \hline
     \textsc{Library}(*) & \textsc{\#Types} & \textsc{\#Clients} & $\textsc{\#Clients}_{obs}$ & \textsc{Category} \\ 
      \hline
    	\javax & 6 & 23,211 & 14,442 & Extension\\
    	\commons & 24 & 2,557 & 2,042 & Utility\\
    	\slf & 38 & 31,752 & 21,398 & Logging\\
    	\junit & 281 & 24,454 & 15,583 & Testing\\
    	\hibernate & 2,746 & 539 & 453 & ORM\\
    	\comio & 112 & 35,000 & 21,959  & Utility\\
       \hline
    \end{tabularx}
    \justify
    (*) For readability, we refer to a library using only its artifactId and version
    \caption{Description of 6 illustrative library examples.}
    \label{tab:six-libs}
\end{table}

\subsection{API usages collection}
\label{sec:static-analysis}

\begin{table}[ht!]
\tiny
\begin{center}
\begin{tabularx}{\textwidth}{rllllll}
    & \multicolumn{3}{c@{\quad}}{\textsc{Libraries Overview}} & \multicolumn{3}{c@{\quad}}{\textsc{Libraries Members}} \\
    
    \cmidrule(r){2-4}\cmidrule(r){5-7}
    
    & \textsc{\#Membs} & \textsc{\#In. Dep.} & \textsc{\#Dist-Clis.} & \textsc{\#Types} & \textsc{\#Meths} & \textsc{\#Fields} \\
    
    \cmidrule(r){1-4}\cmidrule(r){5-7}
  
   \texttt{Min.} & 8 & 0 & 0 & 6 & 1 & 0\\ 
  \texttt{1st Qu.} & 1102.00 & 11 & 6 & 101 & 889.50 & 46\\ 
  \texttt{Median} & 2333 & 56 & 21 & 221 & 1922 & 158\\ 
  \texttt{Mean} & 7895.59 & 479.13 & 89.64 & 662.07 & 6617.91 & 615.61\\ 
  \texttt{3rd Qu.} & 4813 & 261 & 67 & 458 & 4101.50 & 360\\ 
  \texttt{Max.} & 118690 & 47819 & 5375 & 10256 & 108117 & 13682\\  
    
    \cmidrule(r){1-4}\cmidrule(r){5-7}
    
    \texttt{Total} & 41,085,887 & \nbrdependency & 475,928 & 3,453,949 & 34,120,704 & 3,511,234 \\
    \cmidrule(r){1-4}\cmidrule(r){5-7}

    &&&&&& \\

    & \multicolumn{3}{c@{\quad}}{\textsc{Libraries Types}}&\multicolumn{2}{c@{\quad}}{\textsc{Clients Overview}} &  \\
    
    \cmidrule(r){2-4}\cmidrule(r){5-6}
    
    & \textsc{\#Classes}  & \textsc{\#Intfcs.} & \textsc{\#Anns.} & \textsc{\#Type Usgs.} & \textsc{\#Out. Dep.} & \\ 

    \cmidrule(r){1-4}\cmidrule(r){5-6}
    
    \texttt{Min.} & 0 & 0 & 0 & 0 & 1 &\\ 
    \texttt{1st Qu.} & 58 & 8 & 0 & 1 & 1 & \\ 
    \texttt{Median} & 145 & 29 & 0 & 5 & 2 & \\ 
    \texttt{Mean} & 477.09 & 80.41 & 8.72 & 25.14 & 2.70 & \\ 
    \texttt{3rd Qu.} & 310 & 66 & 9 & 23 & 3 & \\ 
    \texttt{Max.} & 9,200 & 930 & 98 & 15,379 & 45 & \\ 
    
    \cmidrule(r){1-4}\cmidrule(r){5-6}
    
    \texttt{Total} & 2,964,707  & 442,157 & 47,085 & 21,268,765 & \nbrdependency & \\
    
    \cmidrule(r){1-4}\cmidrule(r){5-6}
\end{tabularx}
\end{center}
\caption{Descriptive statistics of libraries (GAV) and clients (GAV)}
\label{tab:desc}
\end{table}

We collect the \jar file of each version of each of our \nblibraryga \libs from \mc and statically analyze it to extract all its API members.
Then, we store this list of members in a relational database. 
\autoref{tab:desc} shows descriptive statistics about the APIs and clients for our study. 
The \textsc{Libraries Overview} part shows the number of API members (types, methods, fields) in our set of libraries, the number of dependencies declared towards these libraries and the number of distinct clients that declare these dependencies. 
The \textsc{Libraries Members} part  details  the distribution of the number of type definitions and the total number of  methods and fields across the library APIs.
The \textsc{Libraries Types} part distinguishes between  different kinds of type definitions (classes interfaces and annotations) that we found in APIs. We provide a detailed description of these types since they will form the main granularity at which we analyze API usages.
The smallest API in our dataset is the  \textit{javax.inject:javax.inject:1} library, which contains 1 interface and 5 annotations, of which, only one defines a default method.

\modified{In a second step, we collect, from \mc, the \jar file of every single artifact that declares a dependency to at least one of the libraries in our dataset.} The artifacts are resolved with Eclipse Aether~\cite{Aether2019}, a Java library that fetches artifacts from remote repositories for local consumption. We analyze the bytecode of each of these clients, looking for local variables, fields, parameters, return types, annotations, type extensions or implementations that are referencing library types, including in lambda expressions. We also analyze invocations that target any element of the resolved API members. The bytecode analysis is implemented on top of ASM~\cite{ASM2019}, a popular Java library for bytecode manipulation and analysis. The source code is available on GitHub~\cite{CodeRepo2019}. For each API usage, we count the number of times an element is referenced. The \textsc{Clients Overview} part of \autoref{tab:desc}  gives the distribution of the number API types used, as well as the number of dependencies declared by each client.


\input{figures/usages.tex}

Table~\ref{tab:client-ex} is an excerpt of the database of API usages that we collected. This excerpt corresponds to the usages collected in the bytecode corresponding to the example in Listing~\ref{lst:client-ex}. For example, this excerpt  shows that  the class \textit{ClusterEntrypoint} references the class \textit{Logger} of \slf one time, and calls the method \textit{Logger.info} 6 times.


\section{Results}
\label{sec:results}

\subsection{RQ1 \RQone}
\label{sec:res-RQ-1}

In this research question, we examine the cases where a client declares a dependency towards a library in its \pom file, but its bytecode does include any usage of the library's  API. We measure the extent of the phenomenon and investigates its causes.

\begin{figure}
    \centering
    \includegraphics[width=1\columnwidth]{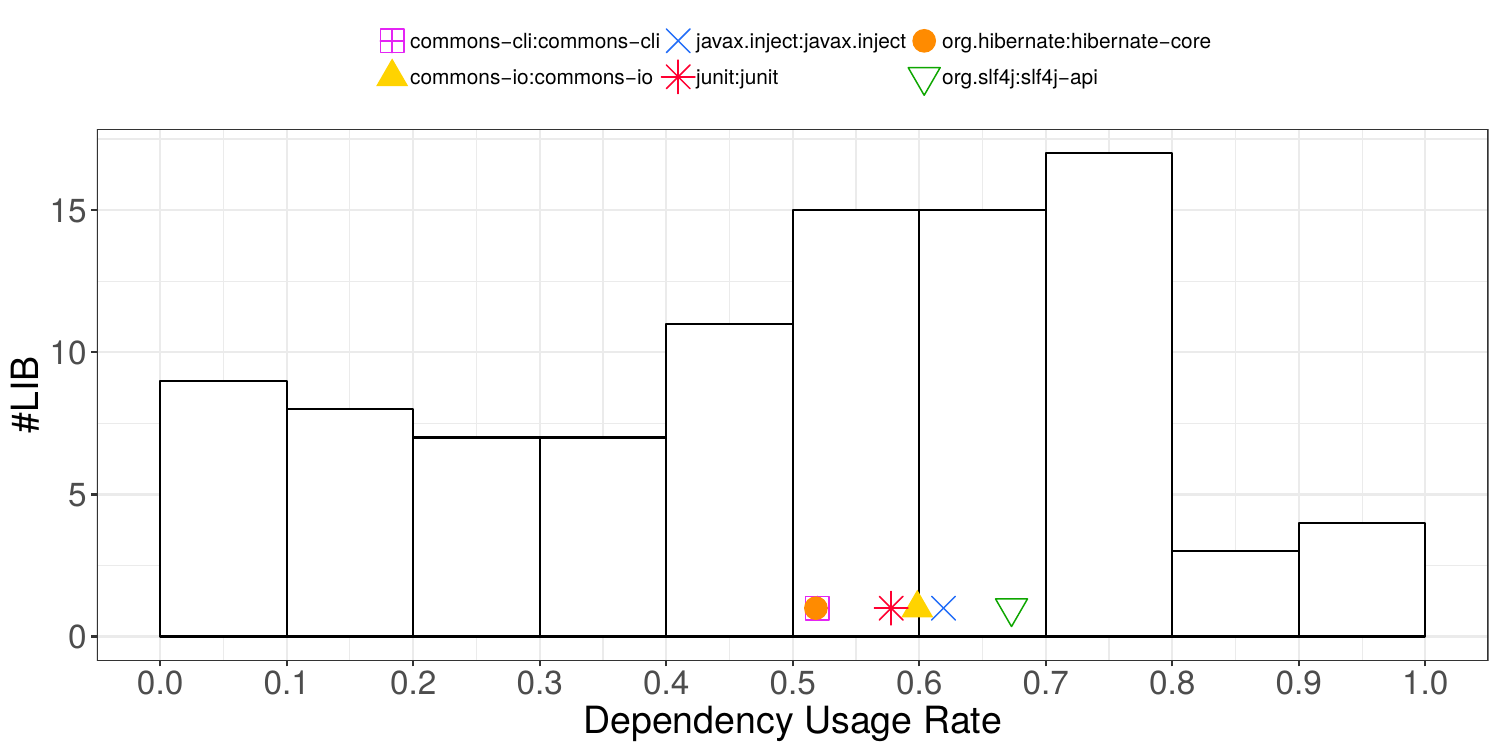}
    \caption{Distribution of dependency usage rate (DUR) among the \nblibraryga \libs. 
    Each bin represents to the number of \libs (y-axis) with a DUR belonging to range of the bin (x-axis).}
    \label{fig:lib_usage}
\end{figure}

\autoref{fig:lib_usage} shows a histogram with the distribution of dependency usage rate $DUR(\mlib)$ among our \nblibraryga \libs. We compute the $DUR$ for every single $library$ in \lib. The leftmost bin includes nine \libs for which less than 10\% of their clients include at least one usage of the \lib's API, \ie with $DUR$ in [0, 0.1[. 
\textit{org.apache.maven:maven-plugin-api} has the  maximum rate, with 96.9\% of its clients that use at least one element of its API. The median rate is 52.4\%. 
No \lib is actually called by 100\% of its clients. 

\begin{sloppypar}
\texttt{spring-\allowbreak{}boot-\allowbreak{}configuration-\allowbreak{}processor} is an example of extremely low $DUR$ ($0.2\%$). This \lib contains a set of annotations, as well as an annotation processor that can be used by IDEs to assist with the development of spring-boot applications\footnote{\url{https://docs.spring.io/spring-boot/docs/2.1.1.RELEASE/reference/htmlsingle/\#configuration-metadata-annotation-processor}}. According to the official documentation, in order to avoid shipping this dependency at runtime, it is recommended to declare it as \texttt{optional}. We suspect that most of the clients that do not mark it as \textit{optional}, do so accidentally. There is however one single client, \texttt{spring-boot-security-saml}{\interfootnotelinepenalty10000\footnote{\url{https://github.com/ulisesbocchio/spring-boot-security-saml}}}, (across its $14$ versions), which does use its API\footnote{\url{https://github.com/ulisesbocchio/spring-boot-security-saml/blob/master/spring-boot-security-saml/src/main/java/com/github/ulisesbocchio/spring/boot/security/saml/util/ConfigPropertiesMarkdownGenerator.java}} to generate Markdown documentation based on the annotations provided by \texttt{spring-\allowbreak{}boot-\allowbreak{}configuration-\allowbreak{}processor}.
\end{sloppypar}

The group of \libs with a $DUR$ below $20\%$, 
is composed of two types of libraries. First, we find libraries that are meant to assist users at development time. For example,  \texttt{spring-boot-configuration-processor} enable a developer to generate customized metadata based on annotations\footnote{\url{https://docs.spring.io/spring-boot/docs/2.1.1.RELEASE/reference/htmlsingle/\#configuration-metadata-annotation-processor}} to provide auto-completion and documentation. Other examples are  \texttt{jaxb-core}, which is used to generate Java source code from XML files, and  \texttt{lombok} that enriches the Java language with annotations that are used to generate boilerplate code. When these libraries are erroneously declared with the default scope (\texttt{compile} instead of \texttt{provided}), they are needlessly considered as a runtime dependency. Second, we distinguish libraries that are not supposed to be called directly by the client. Instead, they are used by other existing dependencies. For example, a client declares \texttt{slf4j} or \texttt{mysql-connector-java} as dependencies in order to let its other dependencies use different logging facades or database connectors.  

\begin{sloppypar}The seven \libs with the highest usage rate among their clients ($DUR(\mlib) > 80\%)$), include the standard libraries for Scala and Kotlin, as well as other frameworks used as domain-specific languages. This latter category includes the  \texttt{maven-plugin-api}, which provides a way for developers to create Maven plugins, and the Apache \texttt{camel-core} library, an integration framework for systems producing and consuming data.
\end{sloppypar}


The majority of \libs, 
$58$ out of the \nblibraryga, have a $DUR$ between $40\%$ and $80\%$. The median $DUR$ of the population is $53.1\%$.
This indicates that, for common libraries of the ecosystem, slightly less than half of clients declare a dependency towards a library and do not make any direct static call to it. For instance, among the $79,364$ clients that declare a dependency toward a version of \textit{commons-io}, only $47,495$ ($59.8\%$) refer to an element of its API in their bytecode. Similarly, the $DUR$ of \textit{slf4j-api}  is $67.3\%$. This corresponds to $117,692$ clients of $174,895$ containing calls to \textit{slf4j-api} in their bytecode.

\emph{Discussing the root cause of unused dependencies:}

We distinguish two common situations where a dependency is declared but not used by a client. First are declared dependencies that ended up in the \pom, most likely, by accident; either through an ingenuous copy-and-paste or inherited from an earlier version of the client's \pom where it was actually used. This hypothesis is consistent with the observations of McIntosh and colleagues~\cite{McIntoshicse2014} who found that build files are more prone to clones than other software artifacts. Take the \textit{javax.inject} for example, which has a dependency usage rate of $61.9\%$, slightly above the median. Since this library contains only 5 annotations and one interface, it is unlikely that any client has used it through reflection. Moreover, the fact that it has only one version (\textit{javax.inject:1}) excludes the hypothesis that this dependency is used to prevent versions conflict. This leaves us with two plausible explanations for the $38.1\%$ of unused dependencies: (1) forgetting to update the \pom and removing unused dependencies during maintenance or (2) a simple copy-and-paste of an existing \pom. 
A living example of the latter hypothesis is the multi-module Maven project \textit{com.eurodyn.qlack2.fuse} where all the modules that declare a dependency to \textit{javax.inject} use at least one API member, except \textit{qlack2-fuse-file-upload-rest}. This module contains only one type~\cite{FileUploadRestTemplate2019} that does not import nor use any member of \textit{javax.inject} API. In this case, it is safe to suggest that this dependency was copied and pasted from another module at the time it was created.

Another common reason for clients to declare a dependency $U$ without actually using it is to expose it in the classpath so that another one of its dependencies $D$ can use $U$. Two mechanisms support this: either $U$ \emph{shadows} another dependency of $D$ and consequently, when the client is built, $D$ uses $U$ instead of the other dependency; or $D$ declares $U$ as an \emph{optional} dependency, which is enforced as soon as the client of $D$ declares $U$ as a dependency.
Classpath shadowing is commonly used by the clients of the \texttt{slf4j} logging framework. \textit{netty}, a framework for building asynchronous network applications, declares  \textit{javassist} as an \textit{optional} dependency,  to accelerate encoding/decoding methods.  \textit{async-http-client}\footnote{\url{https://github.com/AsyncHttpClient/async-http-client/blob/77714b5215afd670d7ca6cd698de21596a0606de/providers/netty4/pom.xml}} depend on \textit{netty} and declares  \textit{javassist} as dependency  to improve \textit{netty}'s performances\footnote{\url{https://github.com/AsyncHttpClient/async-http-client/issues/430}} but no call to \textit{javassist}'s API are present in \textit{asynch-http-client}'s bytecode.

We replicated the study for this research question  with another static analysis tool (based on the Apache \texttt{maven-dependency-analyzer}\footnote{\url{http://maven.apache.org/shared/maven-dependency-analyzer}}), on a  subset of the client - library dependencies \cite{Soto2020}. This study unveiled a similar ratio of unused dependencies:  $44.2\%$ ($19,673$ out of $44,488$) of direct dependencies declared in \pom files were not followed by any static usage of the API of the dependency. This result is consistent with the observation of our current work, i.e., \percdepnotused of declared dependencies do not translate in an actual usage of the dependency. 




\begin{mdframed}[]
\textbf{Answer to RQ1:} 
\nbrdependencynotobs out of \nbrdependency dependencies declared in the clients \pom files are  not used  (\percdepnotused). We observe three main causes: some libraries are not meant to be used directly; developers mistakenly define the scope of the dependency; a \pom file is hardly maintained and cleaned.
These observations indicate the need for better support to analyze and maintain build files. 
\end{mdframed}

\subsection{RQ2: \RQtwo}

In this research question, we study how client usages of APIs are distributed across its types. We focus on the client-library relationships for which we are able to observe at least one usage of the library on the client's bytecode. This represents \nbrdependencyobs dependency relationships between \nbrclientgavobs unique clients, and \nblibrarygavobs unique libraries (\percdepused of the dependencies in our dataset).

\begin{figure}
\begin{subfigure}{\columnwidth}
    \centering
    \caption{Distribution of the share of client using the most used type in the API (Maximum $TUR$).}
    \includegraphics[width=\columnwidth]{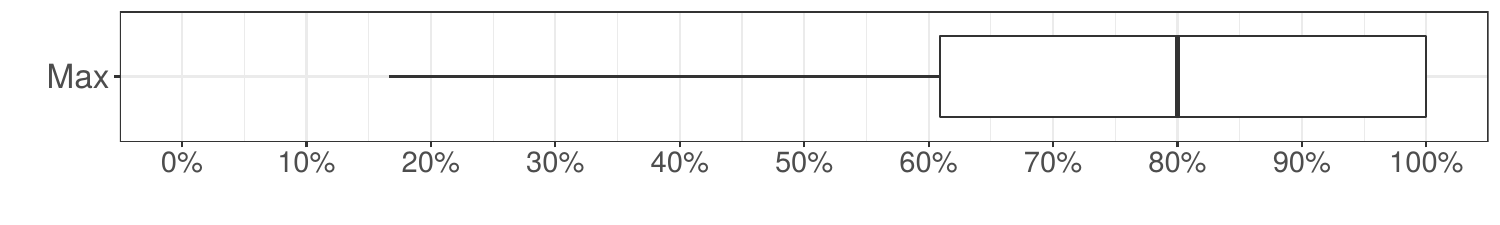}
    \label{fig:most_used_all}
\end{subfigure}
\vspace{-0.6cm}
\newline
\begin{subfigure}{\columnwidth}
    \centering
    \includegraphics[width=\columnwidth]{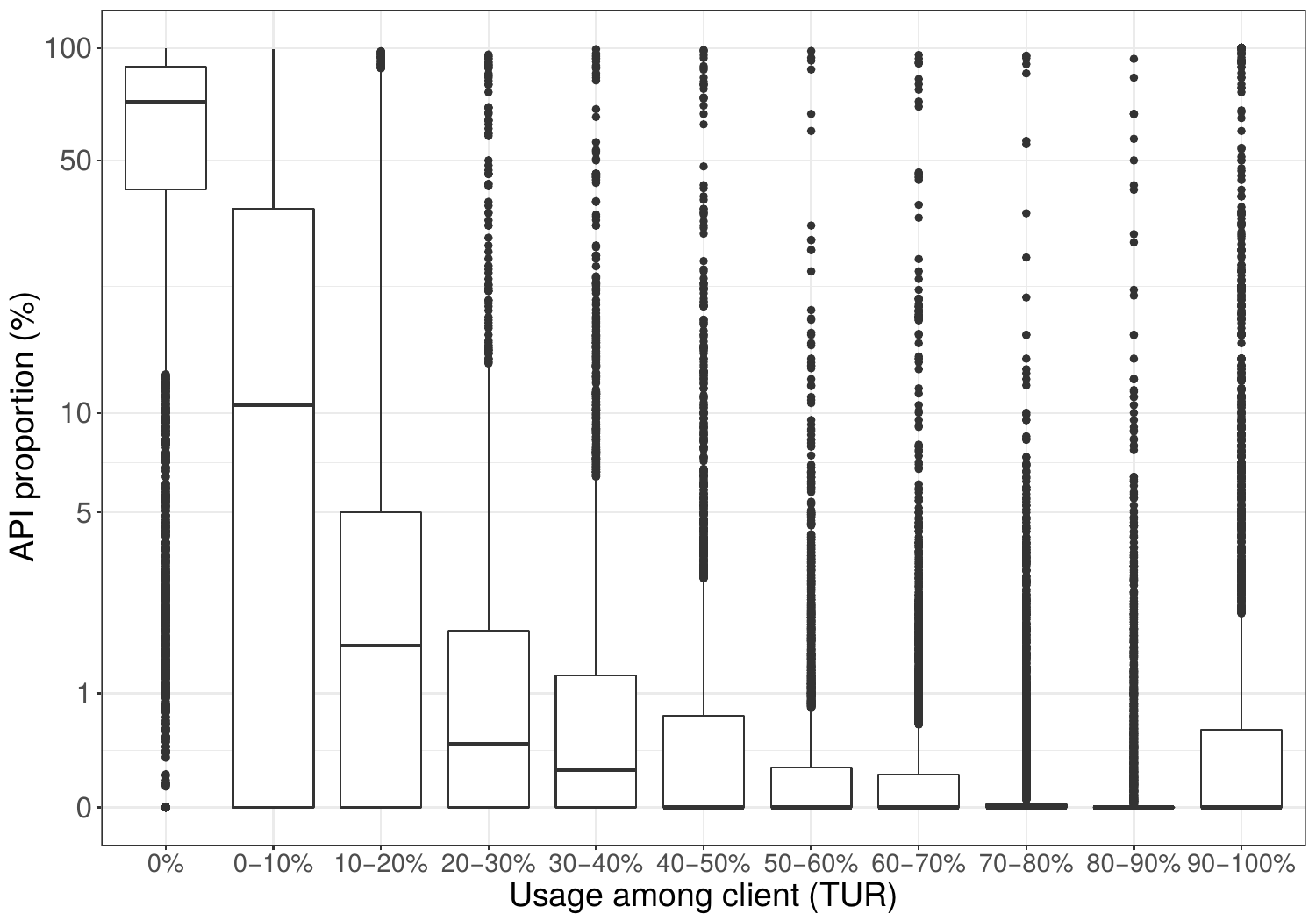}
    \caption{Distribution of type usage rates ($TUR$) of API types}
    \label{fig:class_usage_all}
\end{subfigure}
\caption{Distribution of type usage rates of API types of all \nblibrarygavobs libraries used by at least a client.}
\end{figure}

\autoref{fig:class_usage_all} shows the distribution of usage rates of API types for of all \nblibrarygavobs libraries used by at least a client.
The x-axis represents the Type Usage Rates ($TUR$) grouped in 11 categories. The first category is for types having a $TUR$ that is equal to zero, while the remaining categories are grouped by 10\% ranges, the lowest bound excluded. The y-axis represents the proportion of types in each library that falls into each category, from 0\% to 100\%, on a logarithmic scale. 
This figure is read as follows: The first column on the left represents the distribution of the share of API being used by none of its clients, the second column to the left shows the distribution of the share of API being used by more than 0\% but less than 10\% of its clients, and so forth.

First, let us analyze the share of API types for each library being used by no client (leftmost boxplot). The first quartile of this distribution is $41.7\%$, its median is $71.8\%$, and its third quartile is $88.9\%$. This means that for $50\%$ of libraries, more than $71.8\%$ of the API types are used by no client in our dataset. 

On the opposite side of the figure, the rightmost column shows that the proportion of API types used by more than $90\%$ of clients is greater than $0.6\%$ for $25\%$ of libraries. This hints the existence of a handful number of API types in each library that are used by a vast majority of the clients.
We investigate extreme usages further with \autoref{fig:most_used_all} by showing the distribution of the share of clients using the most used type of each API. The first quartile of this distribution is $60.9\%$, it is median $80.0\%$, and its third quartile is $100\%$. This means that for more than $25\%$ of the libraries there is at least one type used by all clients. Whilst, for more than $75\%$ of them there is at least one API type used by more than $60.9\%$ of the clients.

\begin{figure}
    \centering
    \includegraphics[width=\columnwidth]{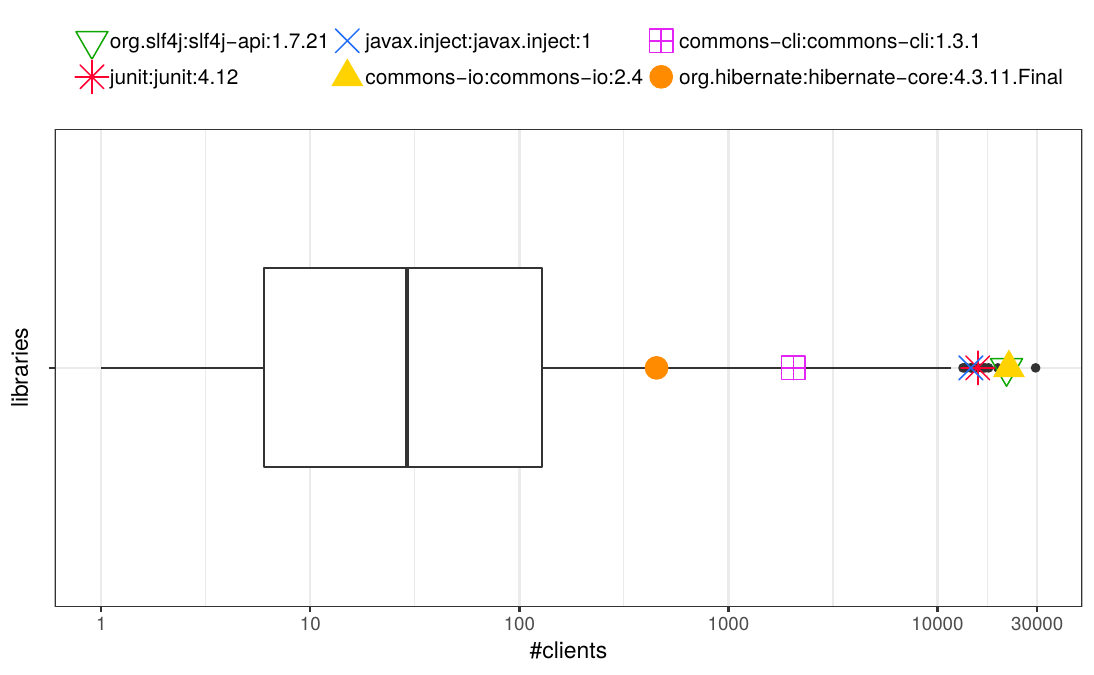}
    \caption{Distribution of the number of clients per library.}
    \label{fig:nb_client_distrib}
\end{figure}

These two observations, a small number of API types used by many clients and a large portion of types used by no clients, are consistent with observations in previous work \cite{Sawant2017,qiu2016understanding}.

Now, let us analyze the second leftmost boxplot. It captures the proportion of API types that are rarely used. These types are used by one client at least, but no more than 10\% of the clients. 
We observe that this category, $TUR \in ]0,10\%[$ exhibits a  large variability: the first quartile is $0\%$, the third quartile is $37\%$, the maximum is $99.4\%$ and the median is $10.5\%$.
This large variability, similar to the leftmost boxplot, suggests two phenomena. First, some libraries have a very large portion of types that are rarely used, a phenomenon that has not been observed previously. Second, these large variabilities might come from large variations in the number of clients for each library. 

\autoref{fig:nb_client_distrib} is a boxplot representing the distribution of the number of clients of each library. The x-axis is the number of clients on a logarithm scale. The plot indicates the quartiles of the distribution. For instance, $365$ of the $4805$ libraries have exactly $1$ client. $75\%$ of libraries have more than $6$ clients, $50\%$ have more than $29$ and $25\%$ have more than $268$. The maximum number of clients observed is $29,466$ for \texttt{scala-library:2.11.8}.


We observe a strong negative correlation ($-0.26$, p-value < $2.2e-16$) between the number of clients that use a library and the share of the API types that are unused. Furthermore, we observe an even stronger positive correlation ($0.34\%$, p-value < $2.2e-16$) between the number of clients using an API and the share of its types that is rarely used ($< 10\%$). This means that libraries with few clients tend to have a large ratio of unused API types. When the number of clients increases, the ratio of unused API types decreases in favor of rarely used types.
These observations hint that the number of library clients matters when studying the API usage. 
The next research question investigates this phenomenon further, with a focus on the most used version of each \lib.






\begin{mdframed}[]
\textbf{Answer to RQ2:} 
All libraries include a handful of API types that are used by a vast majority of their clients.  Meanwhile, $50\%$ of libraries include more than $71.8\%$ of types that are used by no client. These observations confirm previous studies observing that most API clients focus their usage on a small part of the public types. The rate of unused API types varies significantly depending on the number of clients, which motivates a detailed analysis of the most popular library versions. 
\end{mdframed}

\subsection{RQ3: \RQthree}

In this section we focus on the most popular version of each of the $94$ \libs and their \nbmostpopcli clients. The goal is to analyze the distribution of API usages in the presence of a sufficient number of clients. We investigate in particular how the share of API types used by no clients and few client change compared with library with fewer clients.%

%

\begin{figure}
\begin{subfigure}{\columnwidth}
    \centering
    \caption{Distribution of the share of client using the most used type in the most used version of each \lib.}
    \includegraphics[width=\columnwidth]{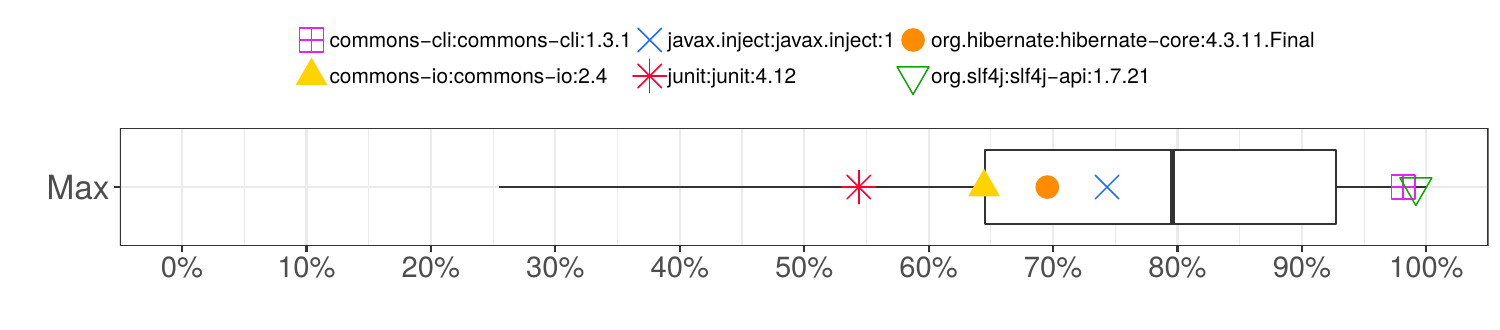}
    \label{fig:most_used}
\end{subfigure}
\vspace{-0.6cm}
\newline
\begin{subfigure}{\columnwidth}
    \centering
    \includegraphics[width=\columnwidth]{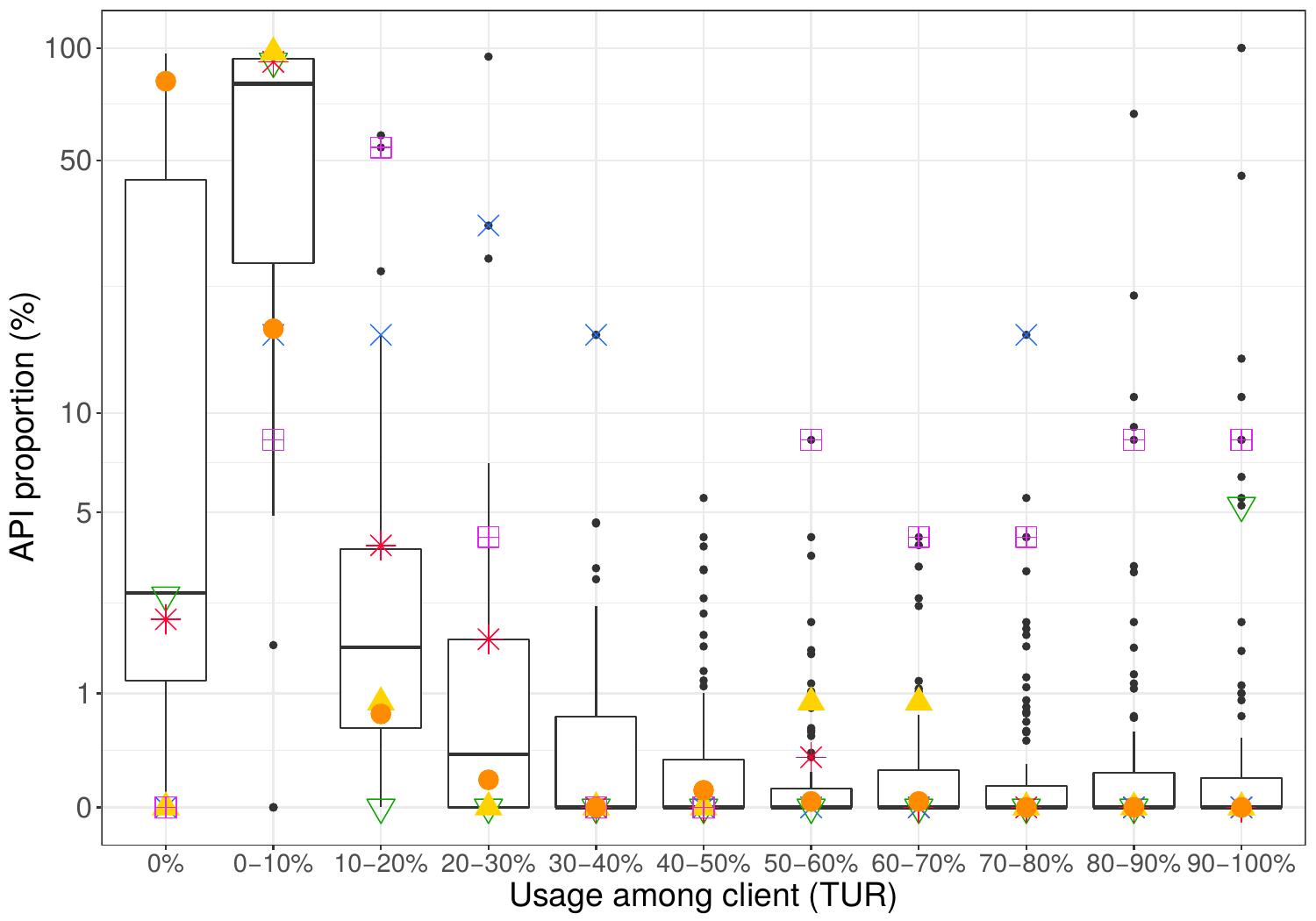}
\caption{Distribution of usage rates of API types of the most used version of each \lib.}
    \label{fig:class_usage}
\end{subfigure}
\caption{Distribution of usage rates of API types of the most used version of each \lib.}
\end{figure}

\autoref{fig:class_usage} shows the distribution of the $TUR$ of API types, focusing on the most used version of each \lib (instead of aggregating all versions as in \autoref{fig:class_usage_all}).The x-axis shows categories based on the percentage of clients using each part of an API. The y-axis represents the share of API types used by a certain ratio of the API clients, on a logarithm scale. The first column shows the distribution of the share of API types of our $94$ libraries used by no client. The second column shows the distribution of the share of API types used by at least one client but less than 10\%.

Overall, we notice that both distributions in figures \ref{fig:class_usage} and \ref{fig:class_usage_all} share the same general characteristic: the proportion of library types  decreases while increasing the $TUR$. Meanwhile, we notice some key differences. First, we remark that the proportion of types used by absolutely no client drastically decreases when focusing on the most popular versions of the \lib, with a median value at 2.6\% (while it was 71.8\% when considering all the versions of the \lib). Second, we observe that the proportion of API types used by less than 10\% of the clients has increased, with a median value of 80.2\%. We observe that with a \textit{sufficient number of users}, for most libraries, the share of API types used by no client falls drastically. 
With a sufficient number of clients, the share of unused API types decreases to the profit of the share of API types that are rarely used.

The third interesting difference while considering the most popular version is about the usage distribution of the most popular type (box plot on top of \autoref{fig:class_usage}): the median does not change, but the quartile values do. 
Precisely, 64\% to 93\%, instead of 61\% to $\sim$100\%. This is consistent with the increase of the quartile values in the categories $[70\%,80\%[$ and $[80\%,90\%[$  and the decrease of the of the quartile values $[90\%,100\%]$. The maximum is usage rate is not 100\% any more, since, with a sufficient number of clients, it is less likely to have all them use the same single type.

Overall, the distribution of the share of clients using the most used type of an API, as well as the share of API types used by more than 50\% of clients, indicates the existence, in most libraries, of a small compact subset of APIs being used by most client. This is consistent with previous work \cite{qiu2016understanding,Lammel2011,Thummalapenta08}.

Here we analyze our illustrative examples in details, and check if our assumption that with enough client all API types end being used by at least one. 
\comio is a good example: it exposes 112 API types, it is used by 21,959 clients and there is no type that is used by no client. We can observe on \autoref{fig:most_used}, that the most used type of \comio is \texttt{IOUtils} used by only $64.5\%$ of its clients. This is lower than 75\% of libraries in our dataset.
Our assumption holds for regular APIs such as \commons, \javax and \slf, which can be partly explained by the small number of types they offer for reuse (resp. 6, 24 and 38).
While our assumption holds, \comio and \slf also have a large share of types that are rarely used (in the $]0,10\%]$ category), indicating a large diversity of usage profiles.

The case of \junit is distinct from the other examples. Our assumption globally holds for this library, since only $6$ of the $281$ public types are not used. The distinctive feature with respect to API usage appears in the boxplot at the top of \autoref{fig:most_used}: the most used type of \junit, (\textit{org.junit.Assert}), is used by only $54.4\%$ of its clients. This singular case can be explained by the fact that version 4.x of junit contains both a new API (including the type \textit{org.junit.Assert}), and the API of version 3.x for backward compatibility reasons (including a type \textit{junit.framework.Assert}). 

\hibernate, our 6th example,  is a counter-example. It exposes 2,746 types, the version we analyze is used by $453$ clients, but $81.8\%$ of its types are never used.

\begin{mdframed}[]
\textbf{Answer to RQ3:} 
Focusing on the most popular version of each \lib, we confirm that, in the Maven ecosystem, with a sufficient number of clients, only a very small share of the API types are never used. Meanwhile, we observe a new phenomenon: a large part of API types are used by less than 10\% of clients (median proportion of types used by less than 10\% of clients is 95,00\%). 
\end{mdframed}

\subsection{RQ4: \RQfour}

Java imposes a design constraint on multi-package libraries: a class member must be publicly visible in order to be used by another class, from another package, inside the library. Yet, once a class is public, it is not possible to limit the visibility boundaries to only the packages of the  library. Once a class is visible beyond its package boundaries, it is accessible to the rest of the world.
Even though, several different conventions can inform a library user that a public type is meant for internal usage, such as naming the package \texttt{internal} or annotating the type as such, non is enforced.
Therefore, one could argue that some types are public only to be used internally by the library itself, which could explain a part of the API types that are not used by the clients.
If this was true, we would observe that the types that are not used by external clients are actually used through internal calls.
Here we investigate this hypothesis and its consequences on the results presented above.

We consider the most popular version of each \lib.  For each library, we distinguish between the types that are used by one client at least and the types that are used by no client. For each category of type, we measure the share that is used through inter-package calls inside the library.

\begin{figure}
    \centering
    \includegraphics[width=\columnwidth]{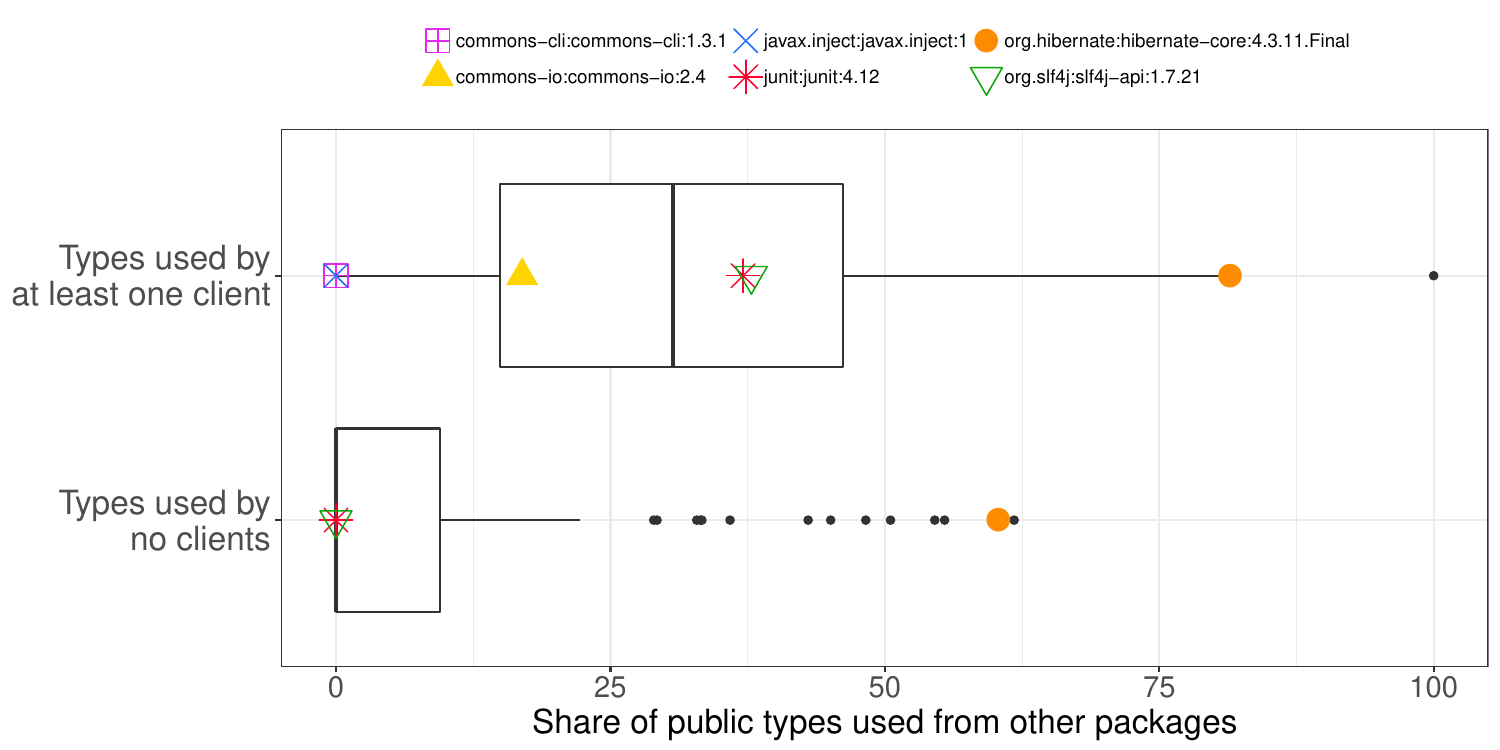}
    \caption{Distribution of inter-package usage rates of API types of the most used version of each \lib.} 
    \label{fig:inter-package_usage}
    \vspace*{-0.3cm}
\end{figure}

The boxplot at the top of \autoref{fig:inter-package_usage} is the distribution of intra-library usages for types that are used by at least one external client. One point on this plot corresponds to the proportion of types of one library that are used by at least one client of this library and that are also used inside the library. 

The boxplot at the bottom of \autoref{fig:inter-package_usage} is the distribution of intra-library usages for types that are used by no client. One point here is the proportion of types of library used by no client but used internally. In this boxplot, $12$ libraries, that have no public type that is used by no clients, do not appear. Among our examples, \commons, and \javax are not on the lower line of the plot since they do not have any unused public type and are single package library.

Furthermore, $9$ libraries have only $1$ package which makes their share of types used from other packages equal to $0\%$ (for both lines). 

The comparison of these two distributions informs us that, not only types used by no clients are not more likely to be used by another package of the same library, but in fact the opposite is true. The median inter-package usage rate for types used by at least a client is $30.7\%$ while it is $0\%$ for types used by no clients at all. Furthermore, a $t-test$ rejects that their mean is the same with \texttt{p-value} $ < 0.001$.
The bottom plot reveals that for more than half of the libraries, no unused type is used by other packages of the library. For all these libraries the types that are publicly visible are not public to allow internal usages, but most probably to be used by the clients of these libraries.
We also observe that 90\% of the libraries have less than 23\% of unused types that are used by another package of the library. This consolidates the observation that API members are not made public for inter-package usages. In other words, types used internally by a library, are not less popular among its clients.
Consequently, \autoref{fig:inter-package_usage} shows that the declaration of some types as public to allow their internal usage, cannot explain the majority of the unused public types.



%


\begin{mdframed}[]
\textbf{Answer to RQ4:} 
The existence of public types that are not used by the clients of a library is not explained by the Java constraint of setting a type as public for internal usages. Based on this new observation and on the results of RQ3, we can conclude that, as soon as a developer sets an element as public, it will likely be used by some client, given a sufficient number of clients, regardless of the developer's  initial intentions.
\end{mdframed}

\subsection{RQ5: \RQfive}


In this last research question we explore how the long tail distribution of API usages can be navigated.
We demonstrate that, if developers are willing to satisfy a majority of their API clients (and not all of them), then they can still identify a small core of API types on which they can focus their efforts and eventually apply the good practices from the literature. 
For example, testing, and documentation efforts can be focused on the small core of API types that are the most used without alienating a large amount of clients. Similarly automated library manipulation such as specialization or automated migration can support only this core while supporting the majority of the population of clients.

The dataset for this question includes the most popular version of each \lib and their clients. The API of each library is reduced to API types  that are used by at least one client.

\begin{figure}[t!]
    \centering
    \includegraphics[width=\columnwidth]{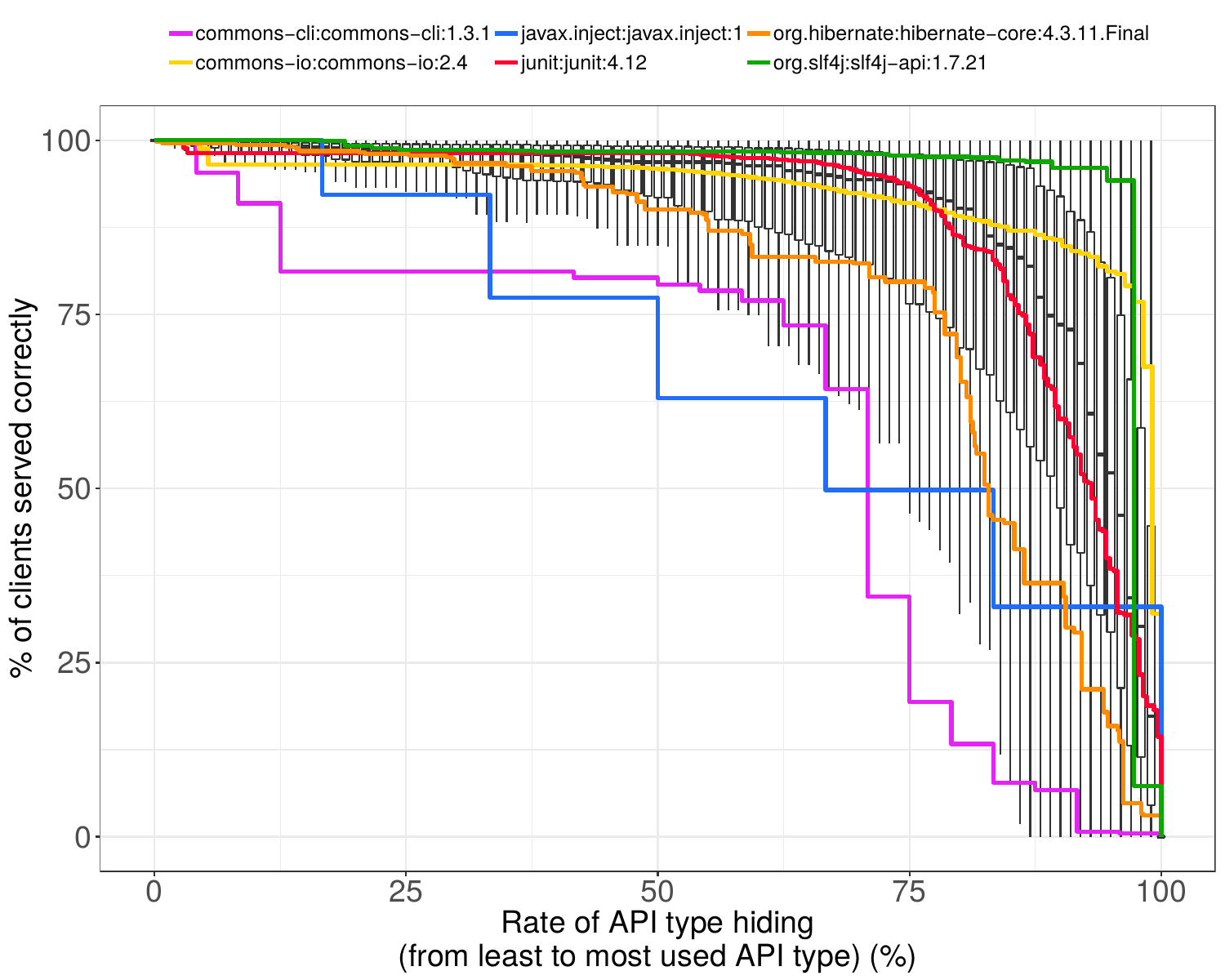}
    \caption{Distribution of extinction sequences for API types. Each sequence simulates the hiding of API types from the least to the most used type. Colored lines show the extinction of 6 libraries, while the boxplot represents the distribution of all $94$ most used version}
    \label{fig:surv-6}
\end{figure}

\begin{figure}[ht]
  \centering
  \includegraphics[width=\columnwidth]{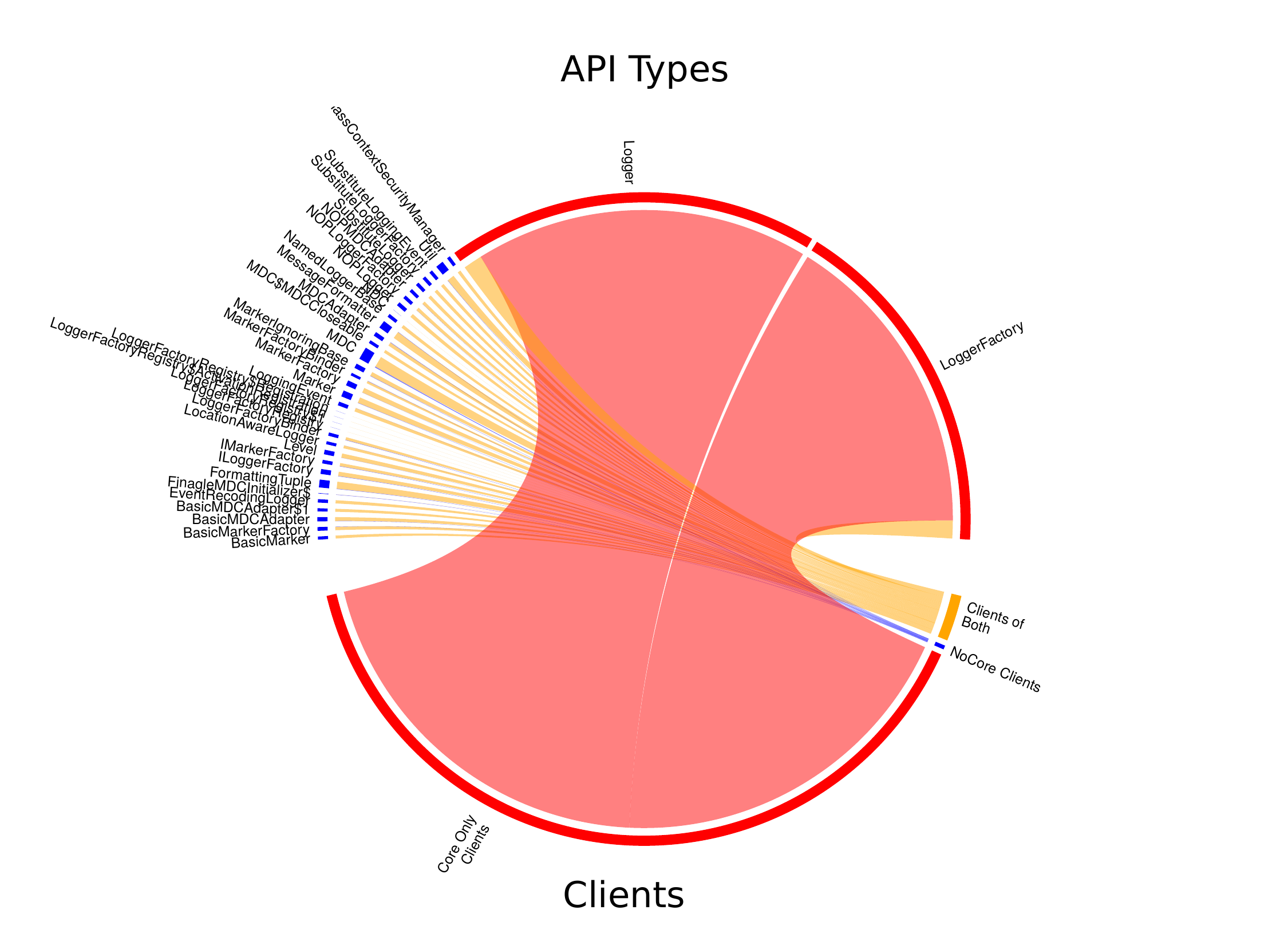}
  \caption{Chord diagram representing the bipartite graph of the \slfgav API types and its clients. Nodes on the upper part represent API Types, with a size proportional to the number of clients using them. The lower part represents three groups of clients (with a proportional size): in red, clients only using the two most popular types (\texttt{Logger} and \texttt{LoggerFactory}), in blue, clients that do not use any of these two types, and in yellow other clients.} 
  \label{fig:slf4j} 
\end{figure}
\autoref{fig:surv-6} represents the distribution of extinction sequences for the $94$ libraries in our dataset. We simulate these extinction sequences by hiding types of their API from the least popular one to the most popular one. At every step of this process we observe the share of client that are no longer able to use the API. The x-axis shows the share of hidden API types (from $0\%$ to $100\%$). Given a share of hidden API types, the y-axis indicates the share of clients that can still access all the types they need. In the rest of this section we say that these clients are correctly served. The colored lines represent the extinction sequences of our $6$ case studies, while the boxplots represent the distributions for the population of $94$ artifacts. All extinction sequences start with 0\% of API types being hidden, and $100\%$ of  clients correctly served. They  all end with $100\%$ of API types hidden and $0\%$ of clients served.

The yellow line represents the extinction sequence for \comio. We observe that it is possible to hide a large part of the types while correctly serving the vast majority of the clients. The big drop occurs when simulating hiding the  $3$ last types of the API: \texttt{FilenameUtils}, \texttt{FileUtils},\texttt{IOUtils}. $14819$ out of the $21959$ clients ($67.5\%$) only use either \texttt{FileUtils}, \texttt{IOUtils} or both, and $7039$ ($32\%$) use only \texttt{IOUtils}, the most used types of the API.

The blue line represents the extinction sequence for \javax. This API has only comports $6$ types ($5$ annotations and an interface) and all of them are used by the clients. Consequently, each simulation of type hiding corresponds to a sharp decrease of correctly served clients. The least popular type is \texttt{Scope}, which is still is used by $1129$ of the $14442$ clients ($7.8\%$). About half of $14442$ clients ($49.8\%$) use only one or both of the most popular types of the API: \texttt{Singleton} and \texttt{inject}.

\commons's extinction sequence is represented in purple. The API of \commons is another example of a rather small API ($24$ types). $64.3\%$ of its client ($1313$ out of $2042$) use the $9$ most used types: we observe a steep drop when removing the ninth most used type (\texttt{OptionGroup}). The rest of the sequence presents an unusual shape because the most popular type is mostly used in conjunction with one other of the popular types. With only $10$ clients using \textbf{only} the most used type \texttt{Options}.

\slf's extinction sequence is represented in green. It is one of the most extreme case in our dataset. The API contains $37$ types, but $20164$ of the $21398$ clients ($94.2\%$) use only two types: \texttt{LoggerFactory} and \texttt{Logger}. As only $1560$ clients use only \texttt{Logger}, the sharpest drop in correctly served clients is caused by the removal of \texttt{LoggerFactory}. This occurs because most clients use these two types in conjunction. The rest of the API provides more advanced logging options that only few clients use. \autoref{fig:slf4j} illustrates this singular situation. This figure shows a chord diagram representing types of the API of \slfgav and its clients. Nodes on the upper par represents API Types with a size proportional to the number of clients using them. The lower part represents three group of clients (with a proportional size): in red, clients only using the two most popular types (\texttt{Logger} and \texttt{LoggerFactory}), in blue, clients that do not use any these two types, and in yellow other clients.

The orange line represents \hibernate's extinction sequence. It simulates the progressive hiding of the $507$ types of its API used by at least one client. This sequence informs us that the $397$ least used types of the API may be hidden without affecting more than $75\%$ of the clients.

The boxplots of \autoref{fig:surv-6} represent the distributions of extinction sequences for all $94$ libraries in our dataset. The median values show that for half of libraries, $88\%$ of the API types or more can be hidden while leaving more than $75\%$ of clients unaffected. 
This illustrates the implication of the long tail distribution of API usages. There is a small set of features used by most client, and the rest is used but rarely.
Similarly, the series first quartiles of these distributions show  that for three quarters of libraries $77\%$ or more of the API types may be hidden without affecting at least $75\%$ of clients.

All APIs include a small set of features that serve a vast majority of the clients. This confirms that API developers who are willing to ignore a minority of clients can indeed focus their maintenance, documentation and development efforts on the small subset of the API that is the most used.

\begin{mdframed}[]
\textbf{Answer to RQ5:} 
With enough clients most API types are used, while most clients can be served successfully with a small subset of API types. In particular, for more than half of the libraries in this study, $88\%$ of the API types or more can be hidden while leaving more than $75\%$ of clients unaffected. If API developers are willing to ignore a minority of clients, they can focus their maintenance, documentation and development effort on the small subset of the API that is the most used. It also open opportunities to automate library migration only supporting a limited part of API targeted while supporting a large majority of clients using it.
\end{mdframed}

\section{Discussion}\label{sec:discussion}
In this section, we reflect on how our observations could hold in other dependency ecosystems. We articulate this reflection around four characteristics of our dataset and how they influence our observations.

\subsection{Source code language for clients} 

All artifacts in our dataset are Java bytecode. Yet, the source code may vary  (Java, Scala, Kotlin, Groovy). We analyzed the clients implemented mostly in Scala, to check if a client's source code language affects the usage of external APIs. We choose Scala because it is the most popular language of the ecosystem, aside from Java, according to the popularity of its standard library.

To assess if clients implemented in Scala use APIs differently than those written purely in Java, we select all the clients that  declare a dependency (excluding test scope) towards any version of the Scala standard library (\textit{org.scala-lang:scala-library}. Let us note that these clients might also use other languages, e.g., Java. This subset of clients represents 31.4\% of the dataset used for RQ2 (212 844 artifacts). 
Then, we select libraries which API is called both by clients using Scala and clients that do not include Scala. This represents 1322 out of the 4931 clients of our dataset. 

We perform a Welch test to determine if the average number of types of the API called by Scala clients is significantly different from the one called by clients not using Scala. 
While the average number of types used by clients using Scala (12.6) is slightly higher than the one of client not using Scala (11.8), we cannot reject the hypothesis that these means are equal (p-value 0.45). 
In other words, clients using Scala do not use significantly more types of an API than other clients.

\subsection{Build tool} 

Our observations about unused dependencies (\autoref{sec:res-RQ-1}) are affected by the build system used to compile and test a client. Indeed, different languages and associated build tools implement different policies regarding external API usages. In RQ1 and in our subsequent studies~\cite{Soto2020} we observed that Maven can build and package projects that include libraries that are not used. Meanwhile, the Go compiler does not compile programs which declare unused imports\footnote{\url{https://golang.org/doc/faq\#unused_variables_and_imports}}, and widely used tools such as goimports\footnote{\url{https://godoc.org/golang.org/x/tools/cmd/goimports}} can refactor imports of a go program to remove unused ones. 

The ecosystem of Go libraries likely behaves differently than the Java ecosystem, regarding our observations in RQ1. We can therefore speculate that the different practices enforced by the build tools impact the distribution of API usages as studied in this work.


\subsection{API size} 

Our dataset shows  a loose correlation (0.21, p-value 0.04) between the \textit{core-index} of a library and the size of its API (number of public types). Consequently, variations in the range of  sizes for a given set of APIs are likely to change the range of core sizes and the core-index. 

In our dataset, the $3$ libraries that are written in Scala (\texttt{scala-library}, \texttt{scala-compiler}, and \texttt{scala-reflect}), have very large APIs in terms of number of bytecode types, in part because the Scala compiler generates types to implement various Scala language constructs. This tends to exacerbate the importance of unused or rarely used types.

The range of API sizes varies in other ecosystems. For example Abdalkareem and colleagues~\cite{trivialPackage} observe the increasing popularity of \textit{trivial packages} in the npm ecosystem, i.e. packages that contain less than 250 LOC. These packages represent $16.8\%$ of the npm population in 2017. Meanwhile, $90\%$ of our  artifacts have more than $147$ accessible elements. This difference in API sizes between npm and Maven influences the observations about the core of APIs: it is very likely that the core of npm libraries is proportionally larger than in Maven. 

In summary, the trends in size of APIs vary in different ecosystems. 
Yet, these variations should not affect the existence of a core set of API members that are used by a majority of clients.


\subsection{Update frequency and interconnection}

The number of clients for a given library in a specific version influences the shape of the extinction sequence and the number of  used API members. The frequency of updates impacts this number of clients: a popular and stable library attracts more and more clients over time, while libraries that update very frequently have a scattered population of clients over multiple versions~\cite{SotoValero2019}. Indeed, it is common for client projects to keep outdated dependencies~\cite{Kula18,bogart16}. Consequently, variations in update frequency have an impact on the shape of the Core.

Decan and colleagues~\cite{Decan2018}  study  $7$ software ecosystems (Cargo, CPAN, CRAN, npm, NuGet, Packagist and RubyGems), describing the variations in update frequencies, as well as how interconnected the ecosystems are. They observe that, while all ecosystems grow over time, some also increase in ratio of dependency over artifact. This  indicates that in these ecosystems, dependencies are more and more interconnected. They also observe that, across all the studied ecosystems, a small number of artifacts concentrate most of the usages by others. Furthermore, this inequality seems to increase over time. 
In RQ2 we show how a large number of clients tends to imply no or few unused part of APIs (confirming Hyrum's law), but does not change the fact that in general a small part of APIs concentrate most usages. Based on the observations of Decan about various update frequencies in different ecosystems, it is likely that the  relative size of APIs  with no observed client, will likely vary in from one ecosystem to another.


\subsection{The notion of Core outside Maven}

While our dataset focuses on artifacts from Maven Central, we are confident that the results would be similar on other Java corpora. We refer to  previous works with other dataset to elaborate on the generalizability of our findings. Qiu and colleagues~\cite{qiu2016understanding} analyzed 5000 projects  mined from GitHub and observed that API usages follow Zipf's law. This is consistent with our results, and implies that extinction sequences would produce similar results on their corpus.

Lammel and colleagues\cite{Lammel2011} perform API usages analyses similar to ours, based on the SourceForge ecosystem. Their dataset includes 6,286 clients for which they mine API usages. Several of their findings align with ours and comfort our results. (i) The $4$ most used external APIs of their dataset are libraries providing XML parsing, Logging and Testing functionalities, and are also in our dataset. (ii) The number of clients using popular libraries follow a similar decreasing exponential (See \autoref{fig:lib-popularity}). (iii) Based on the number of clients in their dataset,  Lammel and colleagues conclude that most APIs are not well covered by usages, which  does not contradict our observations on libraries with a limited number of clients. (iv) In their case studies, they observe rare projects that largely cover APIs and many projects using only a limited subset of APIs. This is consistent with our observations. Furthermore the low cumulative coverage they observe is consistent with our observation that most clients focus their API usages on a small subset of APIs.

When it comes to other software ecosystems, they vary in their custom values and policies~\cite{bogart16}, which may change the exact values obtained. Our study focuses on the Maven ecosystem and JVM-based code artifacts. We acknowledge that key variations in other ecosystems require replications with other data to understand to what extent our observations about the existence of a reuse-core generalize.


\section{Related Work}
\label{sec:related}


Several existing works have  investigated  the usage of APIs in open-source projects and industrial applications. In this section we discuss the related work along the following aspects.

\textbf{API usage in practice. }
Several studies have focused on understanding how developers actually make use of APIs on a daily basis~\cite{Roover2013,Blom2013,Bauer2014}. Some of the motivations include improving API design~\cite{Myers2016} and increasing developers productivity~\cite{Lim1994}. Qiu et al.~\cite{qiu2016understanding} present empirical evidence showing that a considerable proportion of API members are not widely used, i.e., many classes, methods, and fields of popular libraries have never been used. They have found that, on a corpus of 5,000 projects, API usage distribution follows a power law, which is consistent with our findings.
Sawant et al.~\cite{Sawant2017} propose a tool to mine API usages and evaluate it on a dataset of project mined on GitHub using 5 popular Java APIs. They study how the small set features truly used is often introduced in early version of an API.
Nguyen et al.~\cite{pham2016learning} implement a bytecode based analysis tool to learn API usages of Android frameworks. Their approach is intended to automatically generate recommendations for incomplete API usages, and thus reducing API usage errors and improving code quality. While their dataset covers one application domain, in our paper, we analyze clients of libraries serving different domains.
Kula and colleagues~\cite{Kula18} observe that even if dependency usage is common, maintenance operations on dependencies such as keeping them up to date is often not prioritized.
To our knowledge, none of the previous studies has performed on a population as significant as ours, nor proposed the concept of extinction sequences in this context. 

L{\"a}mmel et al.~\cite{Lammel2011} perform a large-scale study on API usage analysis based on AST elements migration. This is the work that is the most closely related to ours. 
Yet it differs in several important aspects. The size and origin of the dataset: we studied a corpus of more than 800,000 clients from Maven Central, i.e. compiled projects. They studied the sources of 6286 projects from SourceForge. We build our dataset by choosing the most popular libraries and then resolve all the clients of those libraries in the ecosystem.
We discuss other topics such as bloated dependencies and propose the use of extinction sequences to describe API usages.

\textbf{API recommendation and comprehension.}
As open-source software projects continuously grow both in quantity and complexity, recent research has paid special attention to understanding these large systems by studying API properties~\cite{Zheng2011}. In particular, API recommendation systems based on usability~\cite{Stylos2008}, diversity~\cite{Mendez2013}, and stability~\cite{Raemaekers2012} have been proposed. Steidl et al.~\cite{Steidl2012} present an approach based on network metrics to retrieve central classes on large software systems. While this approach relies on internal usages (\ie classes within the same projet) to determine the central classes, in our approach, we rely on external usages.
Thummalapenta et al.~\cite{Thummalapenta08} present a tool that detects hotspots and coldspots of eight widely used open source  frameworks. Their tool is integrated as an Eclipse plugin and aim at helping users of APIs to discover their relevant parts.
Duala~\cite{Duala2012} conducted a study about the common questions that programmers ask when working with unfamiliar APIs. Horvath et al.~\cite{horvath2019} mine client usage Apache Beam to study how developers discover functionalities of the API. They observe a long tail distribution of API usages, which is consistent to our observations.
Our work expands the existing knowledge in the area by characterizing the essential API elements based on the clients' usages, which becomes a valuable criterion to reuse functionalities, i.e., following the wisdom of the crowd.\looseness=-1

\textbf{Software dependency ecosystems. }
During the last decade, researchers have investigated the dependency relationships in software packaging ecosystems~\cite{Mancinelli2016,Pashchenko2018,SotoValero2019}. Research efforts focus on the study of library evolution~\cite{Decan2018}, updating behaviors ~\cite{Raemaekers2018} and the security risks~\cite{Zapata2018}. Bogart and colleagues~\cite{bogart16} highlight the different values and customs of different software ecosystems. Raemaekers et al.~\cite{Raemaekers2013} constructed a Maven dataset of 148,253 \jar files for analyzing the evolution of API members based on code metrics. Gabel et al.~\cite{Gabel2010} perform a study on the uniqueness of source code showing that most existing code is reused code.
Unlike previous work, our study focuses on the analysis of API usages to characterize the reuse-core of API types.\looseness=-1

\section{Threats to Validity}\label{sec:threats}

We report about internal, external, and construct threats to the validity of our study.


\textbf{Internal validity.}
This study relies on a very rich and complex network of software artifacts. The complexity is such that we could not completely resolve the artifacts captured in the MDG~\cite{Benelallam2019}.
Indeed, the MDG contains a minority of artifacts, hosted on other repositories than Maven Central. For network reasons, e.g. download limitations, some artifacts could not be resolved. In total, we resolved \nbrclientgav of the \nbclientgav artifact (91.84\%), which corresponds to \nbrdependency dependency relationships (91.78\%). Our analysis covered \nbusage usages of \nbapielementsused different API elements.
We believe that the results obtained with this large set of APIs and clients represent a good approximation of how clients use popular libraries.

\textbf{External validity. }
Our findings might not generalize to all Java APIs. We selected the  \nblibraryga \libs based on their popularity and on the popularity of \mc. We also noticed that these APIs cover a variety of usage domains (e.g., collections, logging, XML parsing). 
As Maven Central is a collection of opensource components\footnote{\url{https://central.sonatype.org/pages/about.html\#what-is-the-central-repository}}, they may not behave as pure end-applications. \modified{All the client artifacts in our dataset are artifacts from this repository.} This means that the exact sets of types included in the $Core_{n}$ of the libraries studied in this work could be different when observing a different set of clients. 
\modified{Yet, both Qiu and colleagues\cite{qiu2016understanding}, and Lammel and colleagues~\cite{Lammel2011}, observe usages distributions that consistent with ours, on a dataset of Java applications mined respectively on GitHub and SourceForge. Consequently, we are confident about the relevance  of our study subjects and the scale of their dependency relationships.}

\textbf{Construct validity. }
The main threat to construct validity is related to the limitations of static analysis, which may fail to capture dynamic calls from the users to some API members. Reflection and libraries handling dependency injection such as spring-boot, or OSGI plugins allow clients to use API members through dynamic calls. 
Reif and colleagues recently studied the impact of Java dynamic features, which are not soundly handled by static analysis, in the context of call-graphs construction~\cite{Reif2019}. While the empirical evidence show that many projects do use reflection, the prevalence of reflection (proportion of methods that do use it) in their call graph is limited. None of the forms of reflection (namely Trivial Reflection, Locally Resolveable Reflection, and Context-sensitive Reflection) is present in more than $0.47\%$ of the methods in their Top50Maven Corpus.
Consequently, the presence of reflection constructs among client libraries has a minimal impact on our observations. 

\modified{\textbf{Reliability. }
The code to query the Maven Dependency Graph, collect both libraries and client artifacts, and analyze the usages, developed for this study may contain bugs. To limit this threat, $3$ researchers were involved in the iteration of development, analysis of the results, and manual review of data points.
We also made our infrastructure publicly available for further replication~\cite{CodeRepo2019}. Finally, in order to advocate for open-science, we made all the data used in this study publicly available online~\cite{Zenodo2019}. }

\section{Conclusion}

In this paper we study the long tail nature of client-API usages. \modified{We perform a systematic empirical analysis of \nbrdependency dependencies that are declared by \nbrclientgav client artifacts towards the \nblibraryga most popular libraries available in \mc.}

A novel result is the  observation that most of the API types are used by one client at least, when considering the most popular version of an API. For more than half the top \nblibraryga in \mc, less than 2\% of types are used by \modified{no clients of the repository at all.} This original result sets an antecedent to further explore Hyrum's law about behavior usage.
It is interesting to note that this new observation does not contradict with the state of the art: our analyzes also confirms that most APIs have a small number of types that are used by the vast majority of their clients. 
For more than half of the API, only 12\% of types are necessary to serve 75\% of the clients. This means that API developers can focus their effort for maintenance and documentation in order to best serve the majority of their clients.

We envision two main threads of future works. 
First, we wish to explore novel ways of designing public Java APIs in order to reduce the number of types exposed to clients. This may be addressed by the feature of  \textit{modules} introduced in Java 9.
\modified{Second, we will investigate the development of adapters between APIs that provide similar features, focusing on the subset of most used API elements. This is motivated by the growing challenges of dependency management and the need to abstract dependencies from their concrete implementation in order to address these challenges \cite{cox2019surviving}.}

\fussy

\bibliographystyle{cas-model2-names}

\bibliography{main}{}


\end{document}

%% file: figures/code.tex
\begin{lstlisting}[basicstyle=\footnotesize\ttfamily, language=java, float, numbers=left, linewidth=0.92\columnwidth, escapeinside={!}{!}, caption={Code snippet of ClusterEntrypoint class in flink-runtime:1.5.1}, label={lst:client-ex}]
// API members of slf4j-API
import org.slf4j.Logger;
import org.slf4j.LoggerFactory;
// API members of findbugs
import javax.annotation.Nonnull;

!\label{line:implements}! public abstract class ClusterEntrypoint implements AutoCloseableAsync, FatalErrorHandler {

    !\label{line:logger}!protected static final Logger LOG = LoggerFactory.getLogger(ClusterEntrypoint.class);
    !\label{line:encaps}!private final Configuration conf;
    private final Thread hook;
    ...
    protected ClusterEntrypoint(Configuration conf){ 
      ...
      hook = SHU.addShutdownHook(...);
    }
    public void startCluster() throws ClusterEntrypointException {
      LOG.info("Starting {}.", getClass().getSimpleName());
      try { sContext.runSecured((Callable<Void>) () -> { runCluster(configuration); ... });
        } 
    }
    !\label{line:rem-start}!@Nonnull
    private Configuration generateClusterConfiguration (Configuration conf) {
        !\label{line:rem-end}!final Configuration result = new Configuration();
        ...
        return result;
    }
    ...}
\end{lstlisting}

%% file: figures/usages.tex
\begin{table}[htb]
\centering
\scriptsize
\caption{The API usages collected in the code of Listing~\ref{lst:client-ex}.}
\begin{tabular}{lcccc}
  \hline
\textsc{Library} & \textsc{Class}			& \textsc{Member signature}	        & \textsc{\#Calls}\\
\hline
slf4j-api	& org.slf4j.LoggerFactory & getLogger(Class;)Logger;		& 1\\ \cline{2-4}
        	&  				& TYPE			& 1\\
        	& org.slf4j.Logger	& info(String;)V		& 6\\
        	& 				& error(String;Throwable;)V 			& 2\\
\hline
jsr305  & javax.annotation.Nonnull	& TYPE		& 1\\ \cline{2-4}
   	& javax.annotation.Nullable	& TYPE		& 2\\ \cline{2-4}
   	& \parbox{2cm}{javax.annotation.concur
\\ \quad rent.GuardedBy} & TYPE & 9\\
\hline
		\end{tabular}
	\label{tab:client-ex}
\end{table}